\documentclass[pof,amsmath,amssymb,preprint]{revtex4-1}

\usepackage{array}
\usepackage{subfigure}

\usepackage[T1]{fontenc}
\usepackage{fourier}
\usepackage[english]{babel}
\usepackage{amsmath,amsfonts,amsthm}
\usepackage{graphicx}
\usepackage{natbib}
\usepackage{bm}
\usepackage{amssymb}
\usepackage{euscript}
\usepackage{esint}
\usepackage{slashed}
\usepackage{booktabs}

\begin{document}
	
\title{Experimental and theoretical study of metal combustion in oxygen flows}
	
\author{Hazem El-Rabii}
\affiliation{%
	Institut Pprime, CNRS (UPR 3346) -- 1 av. Cl\'ement Ader, 86961 Futuroscope Chasseneuil, France
}%
\author{Kirill A. Kazakov}
\affiliation{%
	Department of Theoretical Physics, Physics Faculty, Moscow State
	University, 119991, Moscow, Russian Federation}
\author{Maryse Muller}
\affiliation{%
	Laboratoire PIMM, CNRS/Arts et M\'etiers ParisTech, 151 bd.de l'H\^opital, 75013 Paris, France
}%
	
\begin{abstract}
The effects of oxygen flow speed and pressure on the iron and mild steel combustion are investigated experimentally and theoretically. The studied specimens are vertical cylindrical rods subjected to an axial oxygen flow and ignited at the upper end by laser irradiation. Three main stages of the combustion process have been identified experimentally: (1) Induction period, during which the rod is heated until an intensive metal oxidation begins at its upper end; (2) Static combustion, during which a laminar liquid ``cap'' slowly grows on the upper rod end; and, after the liquid cap detachment from the sample, (3) Dynamic combustion, which is characterized by a rapid metal consumption and turbulent liquid motions. An analytical description of these stages is given. In particular, a model of the dynamic combustion is constructed based on the turbulent oxygen transport through the liquid metal-oxide flow. This model yields a simple expression for the fraction of metal burned in the process, and allows one to calculate the normal propagation speed of the solid metal--liquid interface as a function of the oxygen flow speed and pressure. A comparison of the theory with the experimental results is made.
\end{abstract}
\maketitle

\section{Introduction}

Since the 1950's, combustion of metals has attracted much attention on account of its numerous practical applications. This interest is motivated by two important features of this process, namely, the large heat of combustion and the refractory nature of metal oxides. The research has been focused primarily on the propulsion technologies which use metallic particles as high-energy additives in propellants to increase the specific impulses and combustion stability of solid rocket motors \cite{maggi2012,greatrix2012}. The refractory nature of combustion products, on the other hand, has been exploited to synthesize a wide range of materials, including (nano) powders, intermetallics, composites, and functionally graded materials \cite{merzhanov1995,borisov2002,yetter2009,sytschev2011,sanin2014}. Many important results have been obtained in these directions which are still subjects of ongoing intensive experimental and theoretical studies \cite{law1973,margolis1985,makino2001,meinkohn2004,meinkohn2006,meinkohn2009,khina2010,kuo2012}.

Yet, the large amount of energy released by a burning metal has undesirable consequences and represents a source of significant fire hazards, especially when metals are used in high-temperature and/or high-pressure oxidizing environments such as those prevailing in nuclear plants and oxygen supply systems \cite{rhein1993,clark1974}. The research conducted in metal-fire prevention has mostly consisted in carrying out standard tests that quantify the relative flammability of different metals, that is their relative propensity to sustain combustion of metallic materials of standardized dimensions in oxygen atmospheres \cite{steinberg1989}. Surprisingly enough, no single test has been developed to date that could provide either absolute ignition limits, or consistent relative ratings for all materials. The reason is probably that a comparatively few effort is devoted to this topic of basic research which is barely covered in the literature. Specifically, theoretical attempts to understand and describe the nature of combustion of bulk metallic materials have been hardly undertaken \cite{hirano1983,shabunya2014}, and to the best of our knowledge, the effect of the melt flow hydrodynamics on the burning process has never been addressed, although its importance has been highlighted \cite{steinberg1991,sato1995}.

The present contribution represents an effort to fill this gap. We report an experimental and theoretical investigation of initiation and progression of the metal combustion in an oxygen flow under normal and elevated pressures. The studied specimens are vertical cylindrical rods ignited by heating up their top surfaces with a laser beam. The experiment shows that the combustion process consists of the following essentially different stages. An {\it induction period}, during which the rod is heated up to a temperature at which an intensive metal oxidation begins at its upper end. This step is followed by a {\it static combustion regime}, during which liquid metal melted by the laser radiation slowly oxidizes, and a liquid ``cap'' grows on the upper rod end. Then, after a slow detachment of the liquid cap from the sample end, a {\it dynamic combustion} begins, which is characterized by a rapid metal consumption and much faster liquid motions. This is the main stage of the combustion process, insofar as most of the metal is burned within it. It turns out to be possible to give an analytical description of these stages which goes beyond mere order-of-magnitude estimates, and in most cases provides a satisfactory quantitative account of the results of observations.

The paper is organized as follows. Section 2 describes the experimental setup. In Section 3, we present and discuss our experimental results for pure iron and mild steel samples. The theoretical description is developed and compared to the experimental data in Section 4. Section 5 summarizes our results.

\section{Experimental setup and procedure}
\begin{figure*}
	\centering
	\includegraphics[width=.7\textwidth]{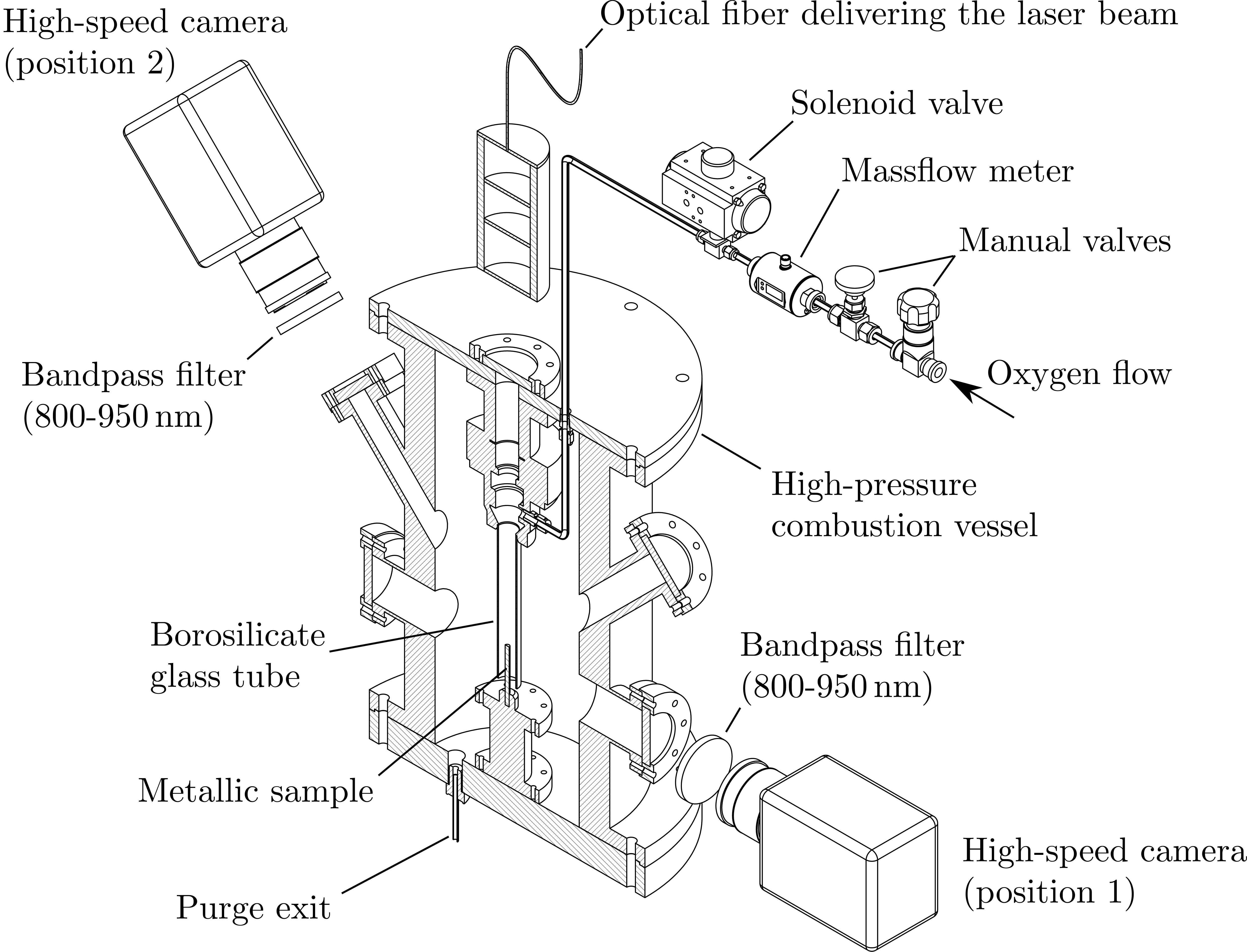}
	\caption{Experimental setup}\label{setup}
\end{figure*}
We studied combustion of 3.0 mm-diameter iron and mild steel rods in an oxygen atmosphere at different pressures and different oxygen flow speeds. The experimental approach provides time-resolved surface temperature, ignition temperature, and high-speed imaging of the heating, ignition, and combustion stages of the metal specimens.

A schematic of the experimental setup used in this study is shown in Fig.~1. The setup includes a high-pressure combustion vessel, a high-speed camera, a mass flow controller, and a continuous fiber laser.

The high-pressure combustion vessel is a 13\,l-capacity, Inconel cylindrical chamber designed to sustain static pressures up to 40 bar and equipped with five silica windows for optical access. Four windows provide visual access (perpendicular and at 45\textdegree, from above, to the cylinder axis), while the fifth, located at the top of the cylinder, lets the laser beam pass into the combustion chamber. The pressure inside the vessel is monitored by means of a pressure sensor.

The material investigated are extra-pure iron (Goodfellow, FE007925, purity 99.99\%) and mild steel (S355J2). The samples are cylindrical rods  15-35\,mm in length and 3\,mm in diameter. The oxidizing atmosphere is supplied by an oxygen gas stream (minimum purity 99.5\%), and is carried into the pressure vessel by a stainless-steel flexible tube before flowing out through the glass tube at a constant flow rate ranging from 0 to 45\,m$^3$/h. The oxygen flow rate is externally adjusted by a manual valve and measured by a thermal mass flow meter.

Ignition is effected by heating up the sample's top surface by a disk laser (TRUMPF TruDisk 10002) operating at 1030 nm. The laser beam is delivered through an optical fiber with a core diameter of 600 \textmu m, providing a uniform intensity distribution. The beam is imaged onto the top of the rod by a set of three lenses (200, 650 and -1000\,mm). The circular beam spot size thus obtained is 3.1 mm in diameter, which ensures a homogeneous heating of the rod surface. Ignition was achieved by varying the laser intensity from 40 to 250 MW m$^{-2}$ (corresponding to the laser power from 320 W to 2 kW) and the pulse duration from 5~ms to 1~s.

The sample burning was visualized using a high-speed camera (Ultima 1024 Photron), with the optical axis perpendicular to the sample axis (position 1), which may operate at a frame rate up to 4\,kHz. Measurement of the top surface temperature of the rod during the ignition and combustion was achieved by recording the heating radiation emitted by the burning sample in the 800-950\,nm wavelength range, with the same high-speed video camera (position 2 - camera axis tilted 45$^\circ$ with respect to the rod axis). Details of the pyrometer calibration are given in \citep{muller2012}. The camera and the laser were triggered by the same signal, ensuring synchronous data acquisition. The timer is set so that the instant $t\,=\,0$ is the beginning of the laser pulse.

The experiment proceeds as follows. The top surface of each sample is first treated with a rough sand paper to ensure sufficient and repeatable absorptivity of the laser radiation. The sample is then housed inside the vessel, where it is fixed at its bottom in a small chuck and partly placed inside a borosilicate glass tube (with an inner diameter of 16 mm). Next, the chamber is purged and filled with oxygen at the desired pressure. After that, an extra oxygen is allowed to stream inside the charged chamber through a glass tube. The rate of the oxygen flow is adjusted by a manual valve so as to reach the desired value of the enclosure pressure. Once ready, the triggering signal is sent.

\section{Experimental results}\label{expresults}

\begin{figure*}
	\centering
	\includegraphics[width=.8\textwidth]{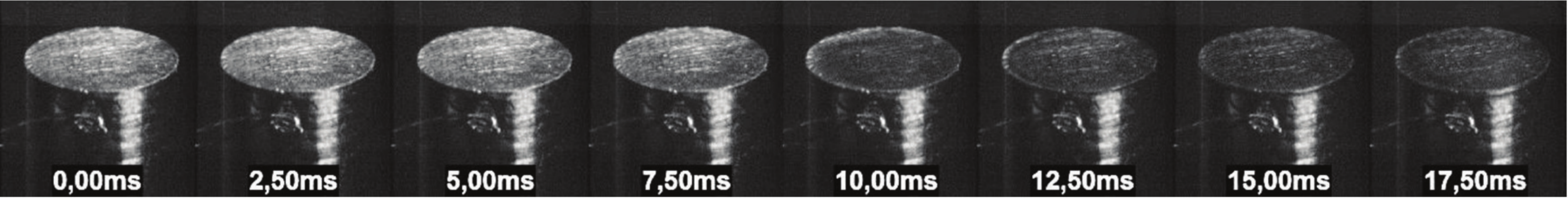}
	\caption{Top surface of a mild steel rod during laser heating at 1\,kW in still oxygen}\label{fig2}
\end{figure*}

\subsection{Ignition and burning in a still oxygen atmosphere}

As a preliminary step, we studied metal combustion in the absence of oxygen flow under normal pressure.

The sequence of images in Fig.~\ref{fig2} shows evolution of the top surface of a mild steel rod exposed to a 1\,kW laser radiation. On the time interval $t\lesssim$ 10\,ms, the surface undergoes a slow solid-state oxidation process at temperatures below 840\,K. Above this temperature, the rate of oxidation significantly increases, as is evidenced by the surface darkening due to a sharp increase of the surface absorptivity from 0.45 to 0.7 \cite{muller2014}. This behavior has been observed for iron and mild steel samples for all laser powers from 180 to 4000\,W, though the temperature at which it occurs increases with the laser power to about 1000\,K at 4\,kW.

\begin{figure*}[t]
	\centering
	\includegraphics[width=.8\textwidth]{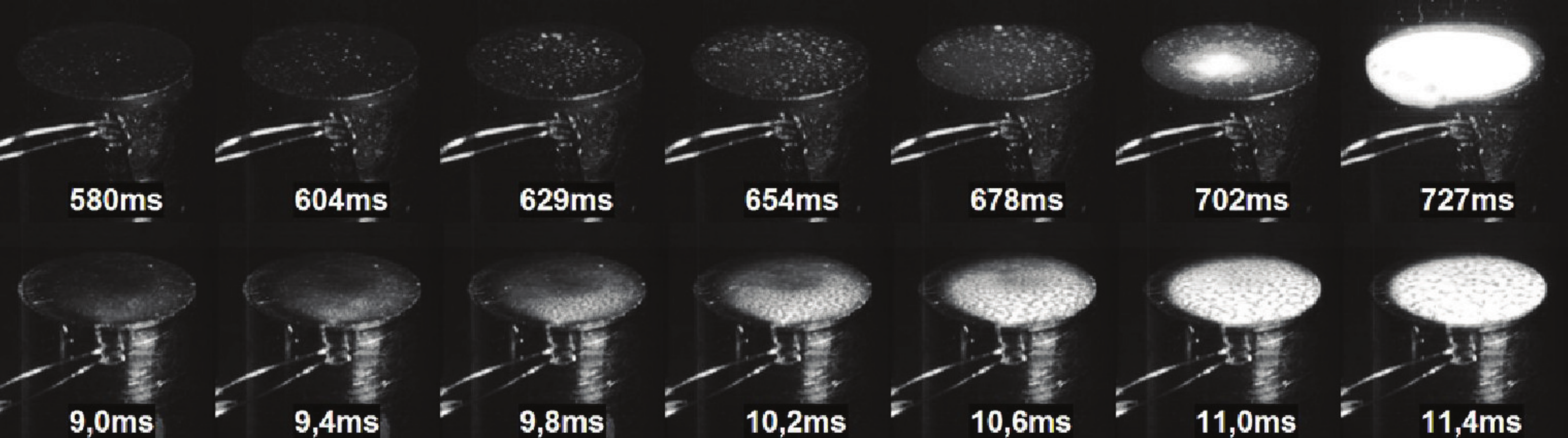}
	\caption{Top surface of a mild steel rod during laser ignition: (a) 180 W, (b) 1.5 kW.}\label{fig3}
\end{figure*}

After this period of slow heating, during which the sample remains in a solid state, a much hotter luminous zone appears on the surface. The sharply increased luminosity clearly indicates the  metal melting, which is also confirmed by the loss of granular structure on the surface (see Fig.\,\ref{fig3}(a) at $t=702$\,ms). For a sufficiently low laser power, Fig.\,\ref{fig3}(a), this area first appears at the center and then extends over the entire surface. At higher laser powers, however, melting occurs uniformly over the surface due to the low surface temperature gradient, Fig.\,\ref{fig3}(b). No ignition was ever observed when the irradiation was interrupted before the onset of melting.

\begin{figure*}
	\centering
	\includegraphics[width=.8\textwidth]{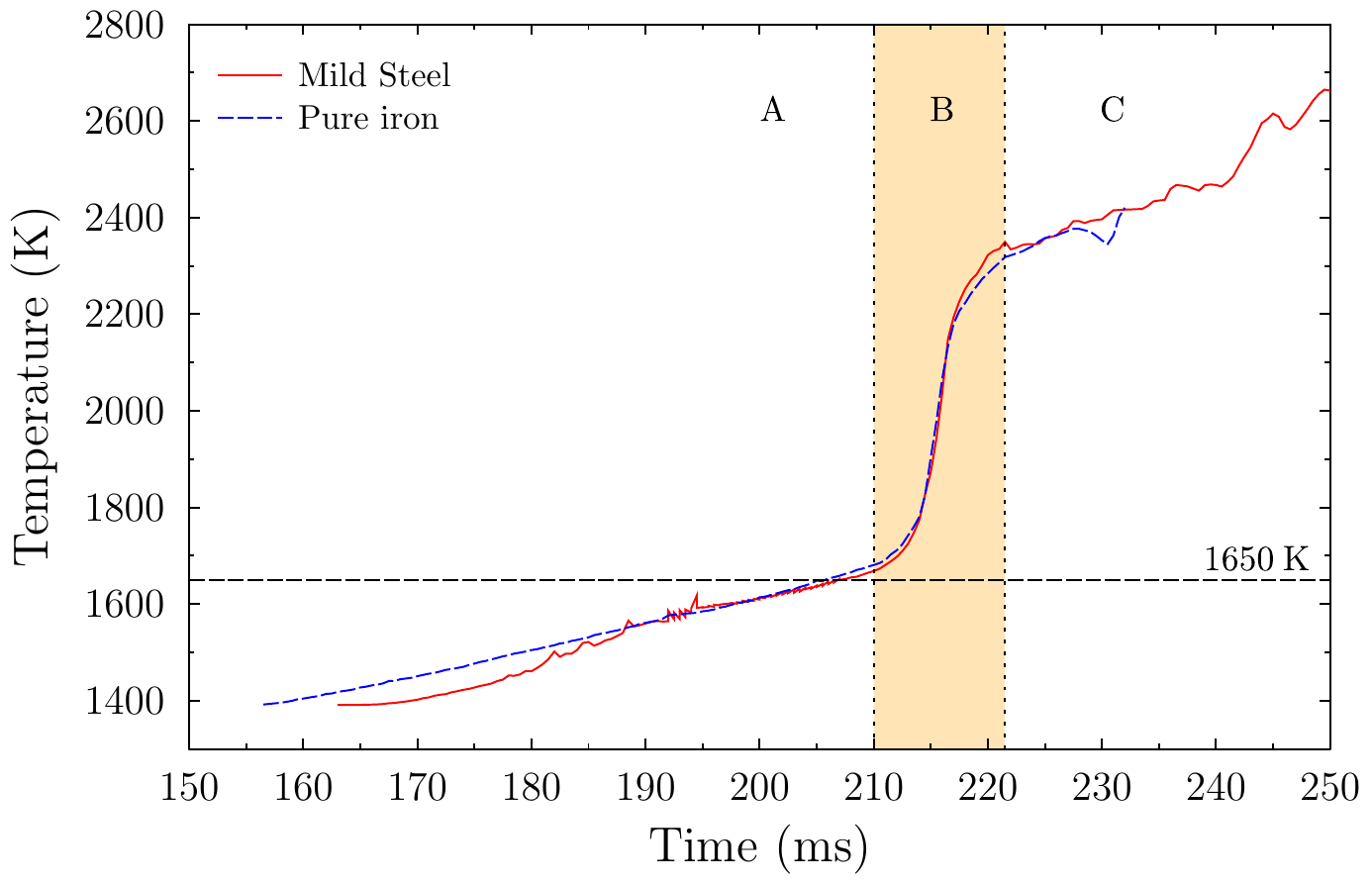}
	\caption{The top surface temperature of an iron and a mild steel rods during laser irradiation at 320\,W in oxygen.}\label{fig4}
\end{figure*}

The melting of the top surface was tracked by the high-speed camera (in position 2) during the laser heating process in oxygen leading to ignition with various laser powers. Emissivity of the sample was assumed to be 0.75 when the temperature exceeds 840 K, in accordance with the absorptivity measurements reported in \cite{muller2014}. Temperature measurements were averaged over a spot of about 0.5 mm in diameter at the center of the top surface.

Figure\,\ref{fig4} shows typical averaged temperature of an iron and a mild steel rods during the last stage of heating by a 320\,W laser. The curves consist of three clearly distinguished pieces. Before ignition, piece (A), the rate of the surface temperature increase is controlled by the laser: it is a resultant of the laser energy input and the energy loss due to the heat conduction into the sample bulk (in the present case this rate is approximately $5\times 10^3$\,K$\cdot$s$^{-1}$). Piece (B) is characterized by a significantly larger slope: the rate sharply rises to about $140\times 10^3$\,K$\cdot$s$^{-1}.$ This means that a new mechanism of heat production comes into play, namely, the iron oxidation. Indeed, the rate change-over occurs at 1650\,K, which corresponds to the melting point of FeO \cite{lide2007}. It is thus explained by an enhanced supply of oxygen to the fresh metal. In fact, the oxygen diffusion coefficient in liquid FeO is several orders of magnitude larger than that in solid FeO \cite{abbaschian2009}, a thin layer of which covers the metal and grows during the heating stage prior to ignition. Our experiments show that this layer retains its protective properties up to 1650\,K for all laser powers in the range 180\,W to 2\,kW. Furthermore, we observed a similar behavior of iron rods under elevated oxygen pressures (10 and 20\,bar). This is as it should be according to the above explanation, for the melting point of FeO changes insignificantly with pressure \cite{shen1993}.

But shortly after the sharp rise, the heating rate drops down to approximately the same value as in (A), which happens at around 2300\,K. This means that above this temperature, on piece (C), the heat released by iron oxidation is spent almost completely on the iron melting. The iron melting point itself, $T_m = 1811$\,K, is not distinguished in any way on the plots of Fig.~\ref{fig4}, because of the latent heat. The heat of iron fusion is $Q_m = 13.8$\,kJ/mol, while its heat capacity $c_p = 25.1$\,J/(mol$\cdot$K), so that the iron melting fully establishes only at a temperature $T_a = T_m + Q_m/c_p \approx 2360$\,K, and then begins to spread down the rod. As the result, a liquid drop grows at the upper rod end, while its apparent temperature (that is, temperature of the drop top surface) continues to rise, though in a much less regular way, because of the more complicated heat transfer in the moving melt. It can reach values as high as 2900\, K, depending on the laser power and the exposure time.

\setlength{\tabcolsep}{2pt}
\begin{figure*}[htpb]
\centering
\begin{tabular}{>{\centering} m{0.5cm} >{\centering} m{15.5cm}}
\footnotesize (a) & \includegraphics[width=15.5cm]{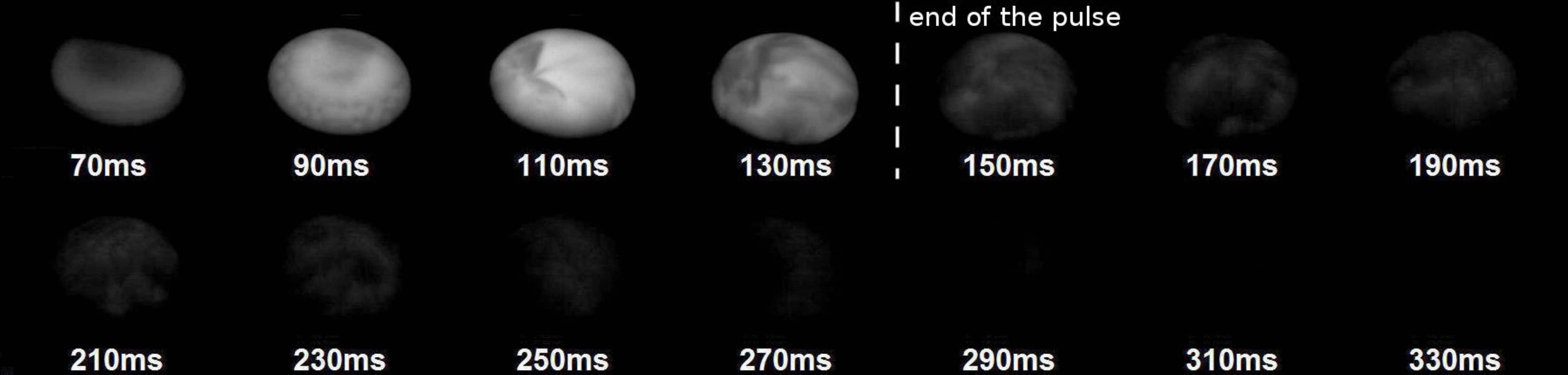}  \tabularnewline \footnotesize (b) &  \includegraphics[width=15.5cm]{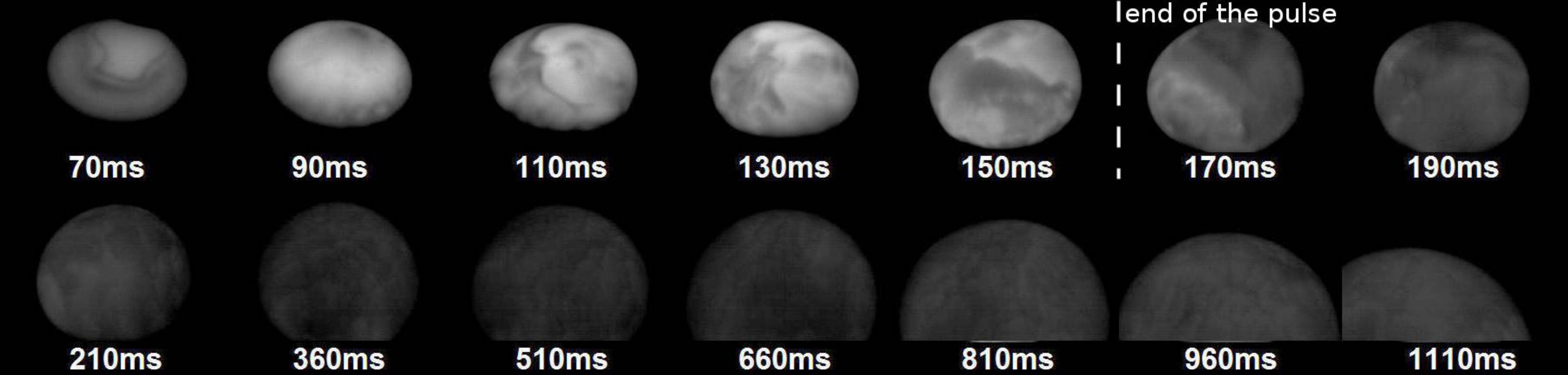} \tabularnewline
\end{tabular}
\caption{Infrared images (800-950 nm) of a pure iron rod while burning in the static combustion regime (640W) for pulse durations of : (a) 130 ms leading to extinction; (b) 150 ms leading to self-sustained combustion, and then to propagation (modified contrast).} \label{fig5}
\end{figure*}
\setlength{\tabcolsep}{6pt}

Once the laser ceases, only the heat from the oxidation reaction is transferred to the rod. If it is not enough to melt the fresh metal, the burning liquid gradually cools down and eventually extinguishes. Otherwise, the combustion enters in a self-sustained regime when the fresh molten metal is delivered into the burning drop until it detaches from the rod, leaving a small amount of molten material on its top. In this case, the liquid temperature decreases and settles at approximately 2400\,K. The cycle of the drop growth and its detachment then repeats itself until the sample is consumed. Time-resolved images of both situations are shown in Fig.\,\ref{fig5}, for pure iron and the laser power of 640\,W. A comprehensive description of this process using the phase diagram of the iron-oxygen system is given in \cite{muller2015}.

\subsection{Burning in an oxygen flow. Normal pressure}\label{flow}
%
\begin{figure*}[htbp]
	\centering
	\subfigure[Oxygen flow speed : 1.7\,m$\cdot$s$^{-1}$, laser power : 640\,W, laser pulse duration : 235\,ms]{\includegraphics[width=.8\textwidth]{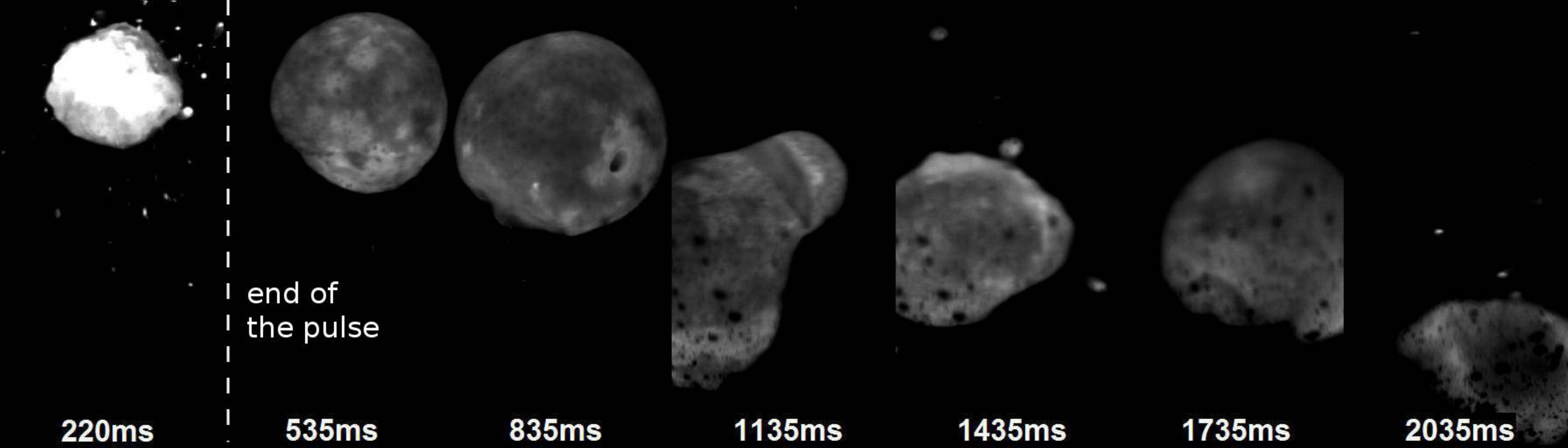}}
	\subfigure[Oxygen flow speed : 1.7\,m$\cdot$s$^{-1}$, laser power : 4\,kW, laser pulse duration : 65\,m.]{  \includegraphics[width=.8\textwidth]{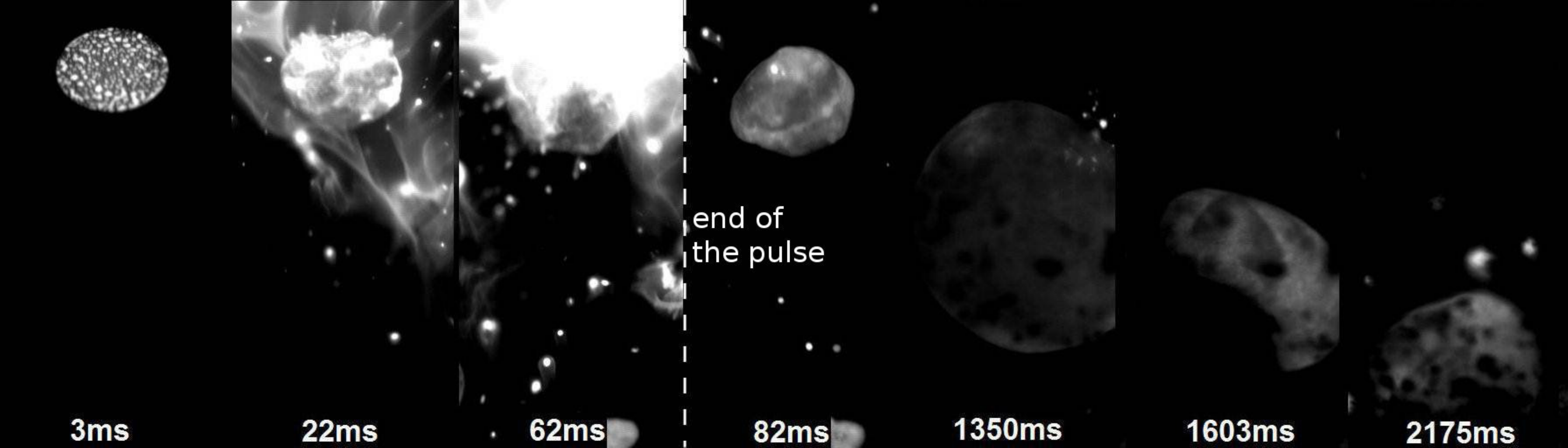}}
	\subfigure[Oxygen flow speed : 10\,m$\cdot$s$^{-1}$, laser power : 180\,W, laser pulse duration :  950\,ms]{\includegraphics[width=.8\textwidth]{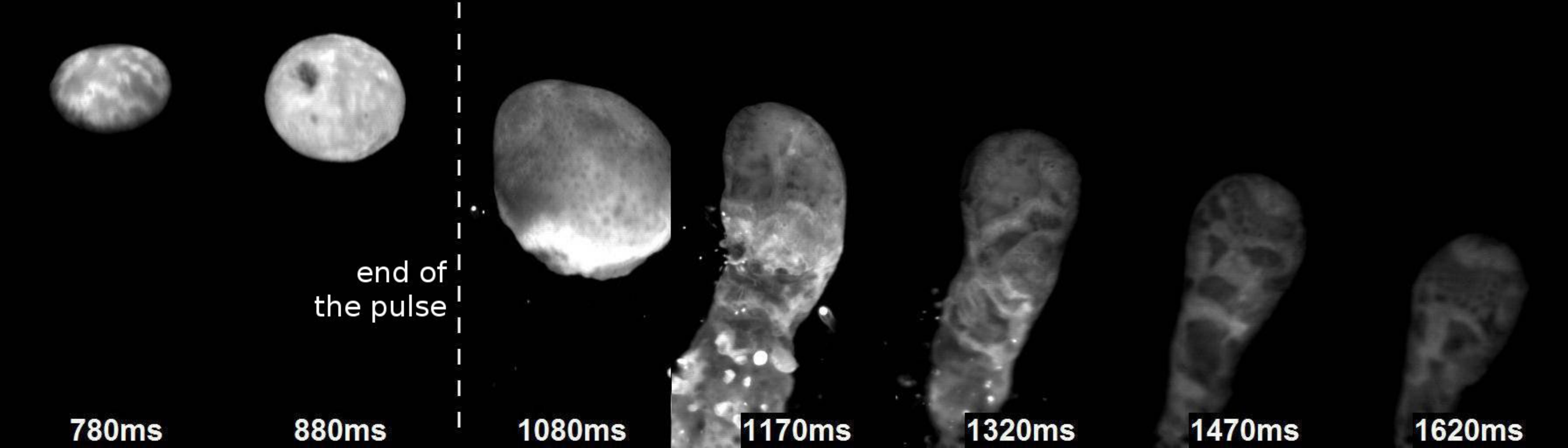}}
	\subfigure[Oxygen flow speed : 10\,m$\cdot$s$^{-1}$, laser power : 2\,kW, laser pulse duration : 60\,ms]{  \includegraphics[width=.8\textwidth]{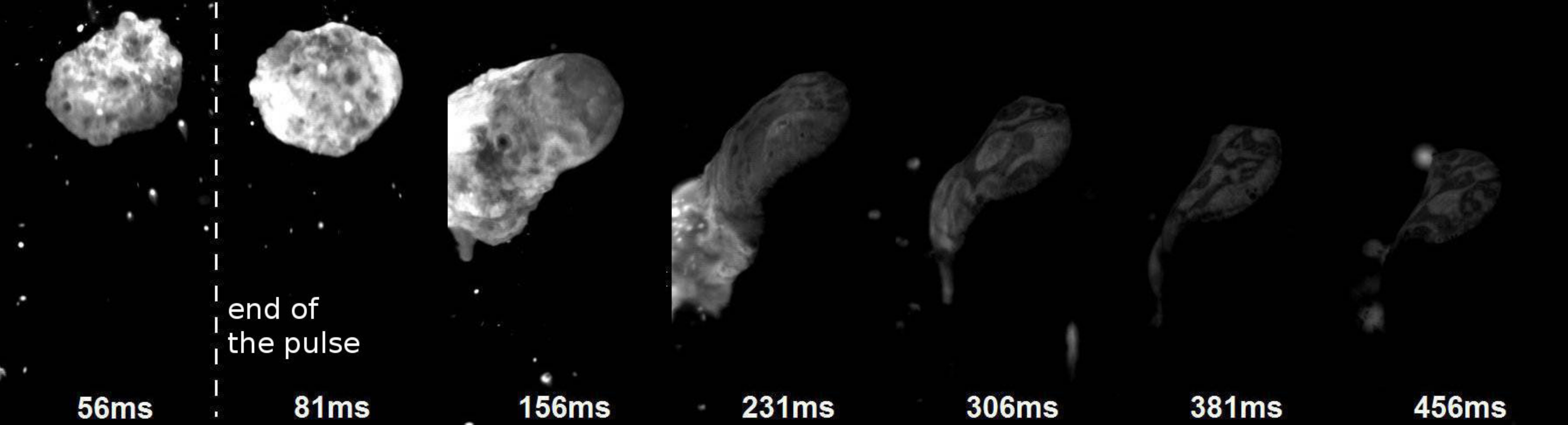}}
	\caption{Infrared images (800-950 nm) of the propagation of combustion on mild steel rods for different oxygen flow speeds and different laser powers. The laser input energies are at the propagation thresholds.} \label{fig6}
\end{figure*}
\setlength{\tabcolsep}{6pt}

\setlength{\tabcolsep}{2pt}
\begin{figure*}[htbp]
\centering
\subfigure[Oxygen flow speed : 30\,m$\cdot$s$^{-1}$, laser power : 1\,kW, laser pulse duration : 50\,ms]{ \includegraphics[width=.8\textwidth]{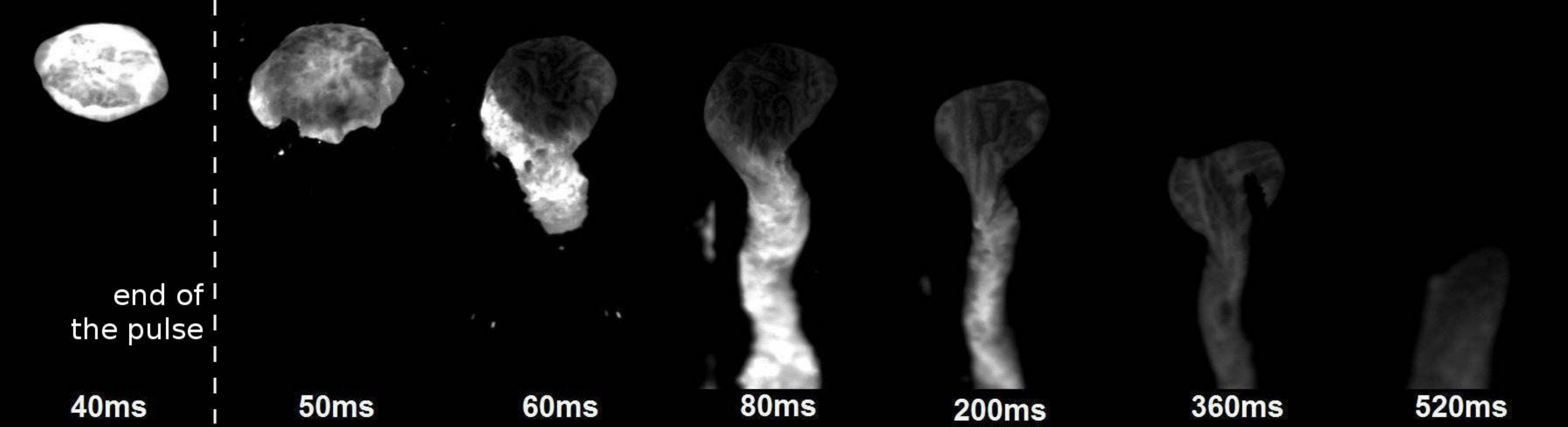}}
\subfigure[Oxygen flow speed : 30\,m$\cdot$s$^{-1}$, laser power : 3\,kW, laser pulse duration : 15\,ms]{ \includegraphics[width=.8\textwidth]{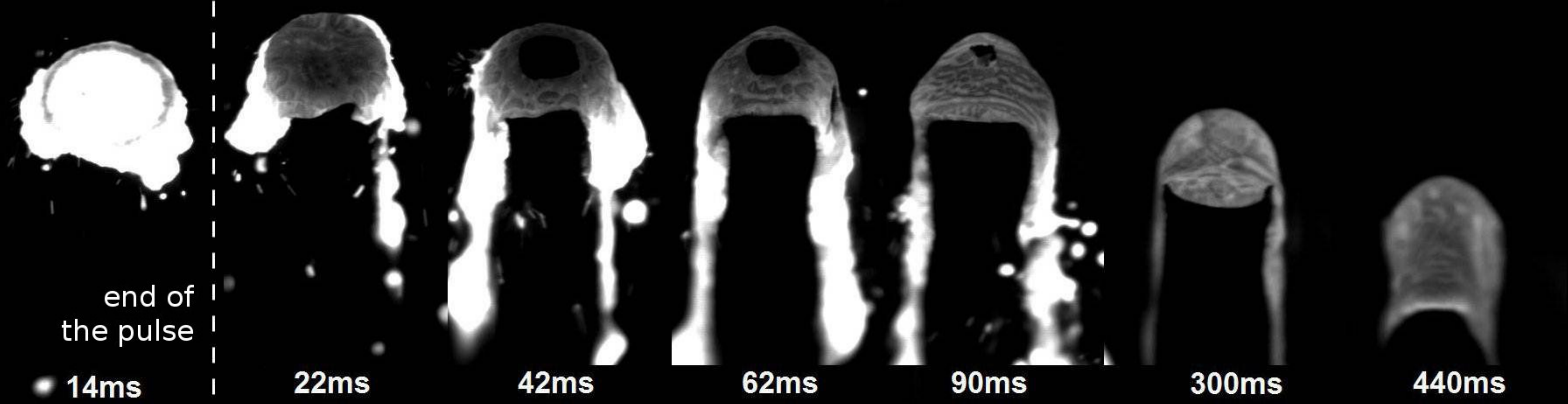}}
\subfigure[Oxygen flow speed : 40\,m$\cdot$s$^{-1}$, laser power : 640\,W, laser pulse duration : 72\,ms]{  \includegraphics[width=.8\textwidth]{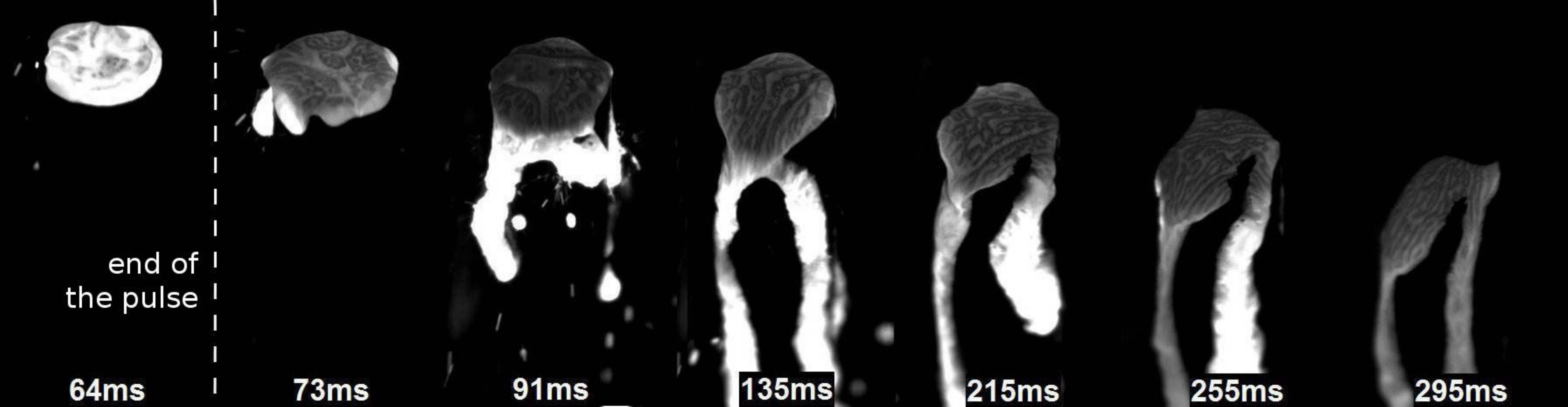}}
\subfigure[Oxygen flow speed : 50\,m$\cdot$s$^{-1}$, laser power : 4\,kW, laser pulse duration : 12\,ms]{\includegraphics[width=.8\textwidth]{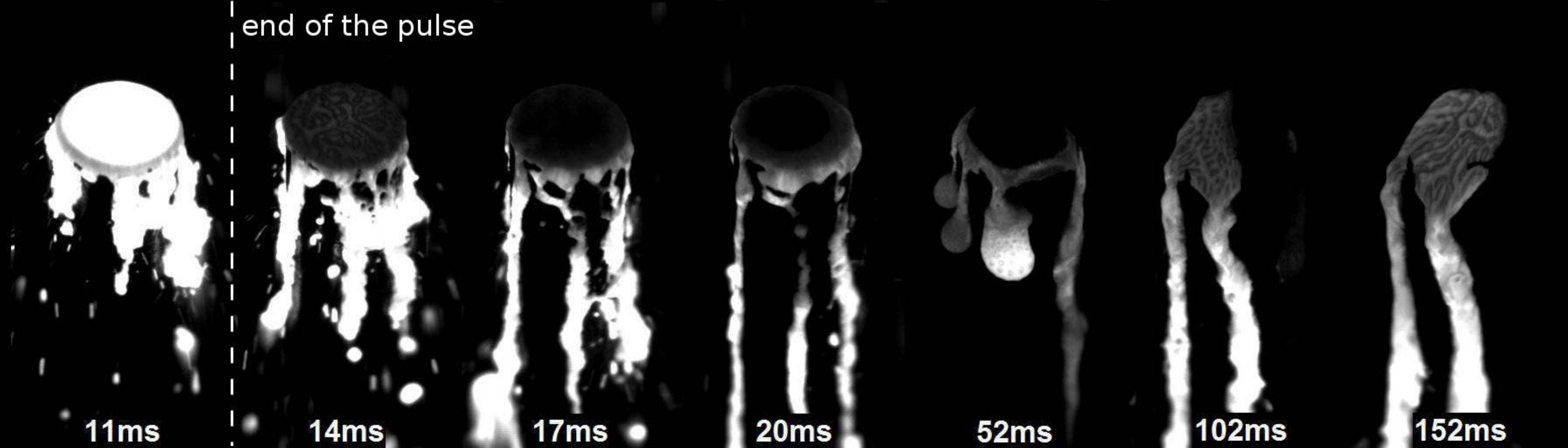}}
\caption{Same as in Fig.~\ref{fig6}.} \label{fig7}
\end{figure*}
\setlength{\tabcolsep}{6pt}

\setlength{\tabcolsep}{2pt}
\begin{figure*}[h]
\centering
\subfigure[Oxygen flow speed : 30\,m$\cdot$s$^{-1}$, laser power : 180\,W, laser pulse duration : 780\,ms]{ \includegraphics[width=.8\textwidth]{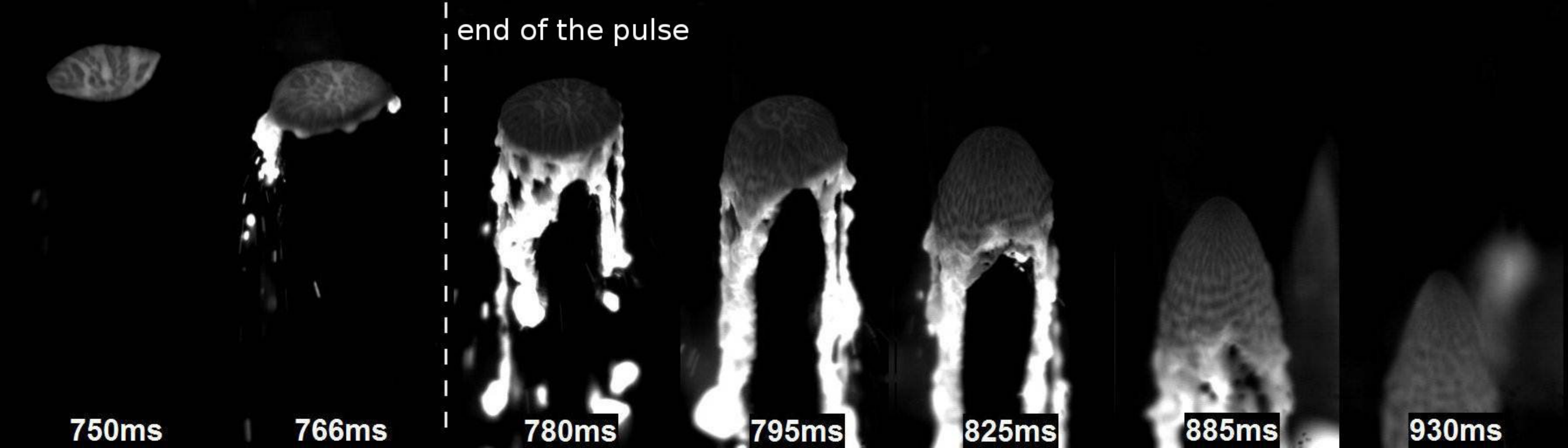}}
\subfigure[Oxygen flow speed : 60\,m$\cdot$s$^{-1}$, laser power : 640\,W, laser pulse duration : 65\,ms]{ \includegraphics[width=.8\textwidth]{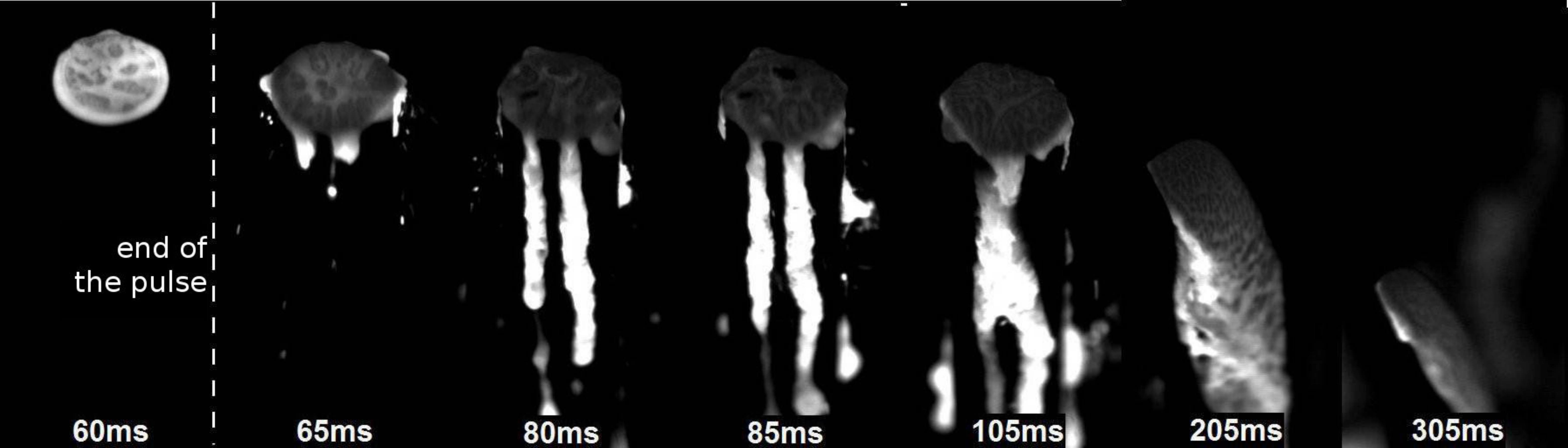}}
\caption{Same as in Fig.~\ref{fig6}.} \label{fig8}
\end{figure*}
We now turn to the combustion of metal rods subject to an oxygen flow. Our aim is to study the effects of the oxygen flow speed and its pressure.

Figures\,\ref{fig6}-\ref{fig8} are sequences of images of mild steel rods burning in oxygen flowing at speeds from $v_0=1.7$\,m$\cdot$s$^{-1}$ to $v_0=60$\,m$\cdot$s$^{-1}$ and various laser powers. In all tests performed, the combustion process ultimately evolved into one of two different self-sustained regimes. At the low flow speeds $v_0\lesssim$1.7\,m$\cdot$s$^{-1}$, Figs.\,\ref{fig6}(a) and (b), the rod burning is quite similar to that observed in the experiments with no oxygen flow, that is, consumption of the rod proceeds through the cycle of growth and detachment of a liquid cap. The detached molten material was often observed to precess around the burning rod as it was running down along it. A different behavior was observed for $v_0\gtrsim$2\,m$\cdot$s$^{-1}.$ Figures\,\ref{fig6}(c),(d) and Figs.\, \ref{fig7}, \ref{fig8} show that in this case, the liquid continues to flow down after the drop fall, which leads to a gradual erosion of the rod on one of its sides with an increasingly steep slope. It never progresses in a symmetrical way with respect to the rod axis.

\begin{figure*}
	\centering
	\subfigure[Oxygen flow speed : 10\,m$\cdot$s$^{-1}$]{\includegraphics[width=.45\textwidth]{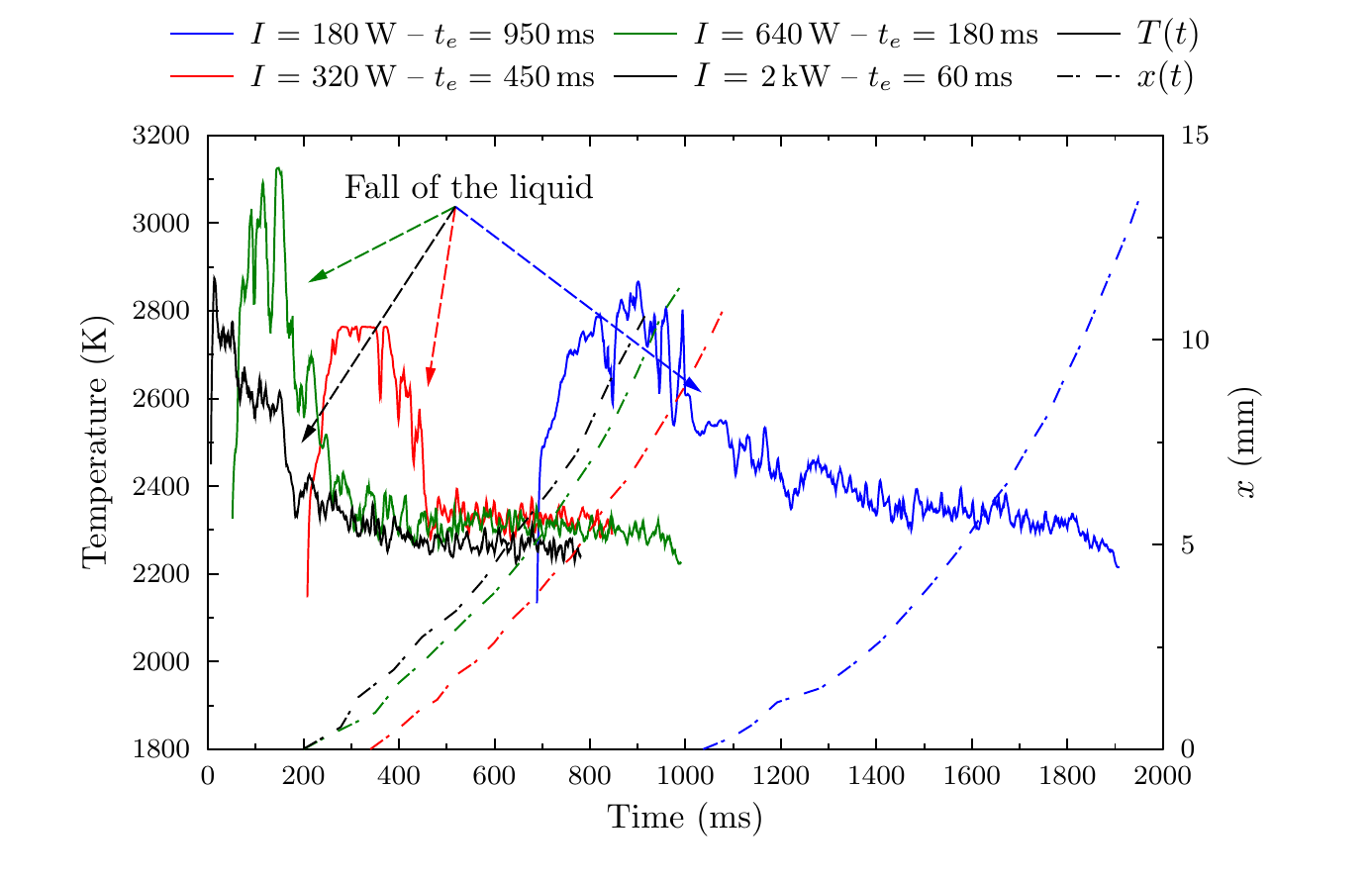}}
	\subfigure[Oxygen flow speed : 10\,m$\cdot$s$^{-1}$, shifted time]{\includegraphics[width=.45\textwidth]{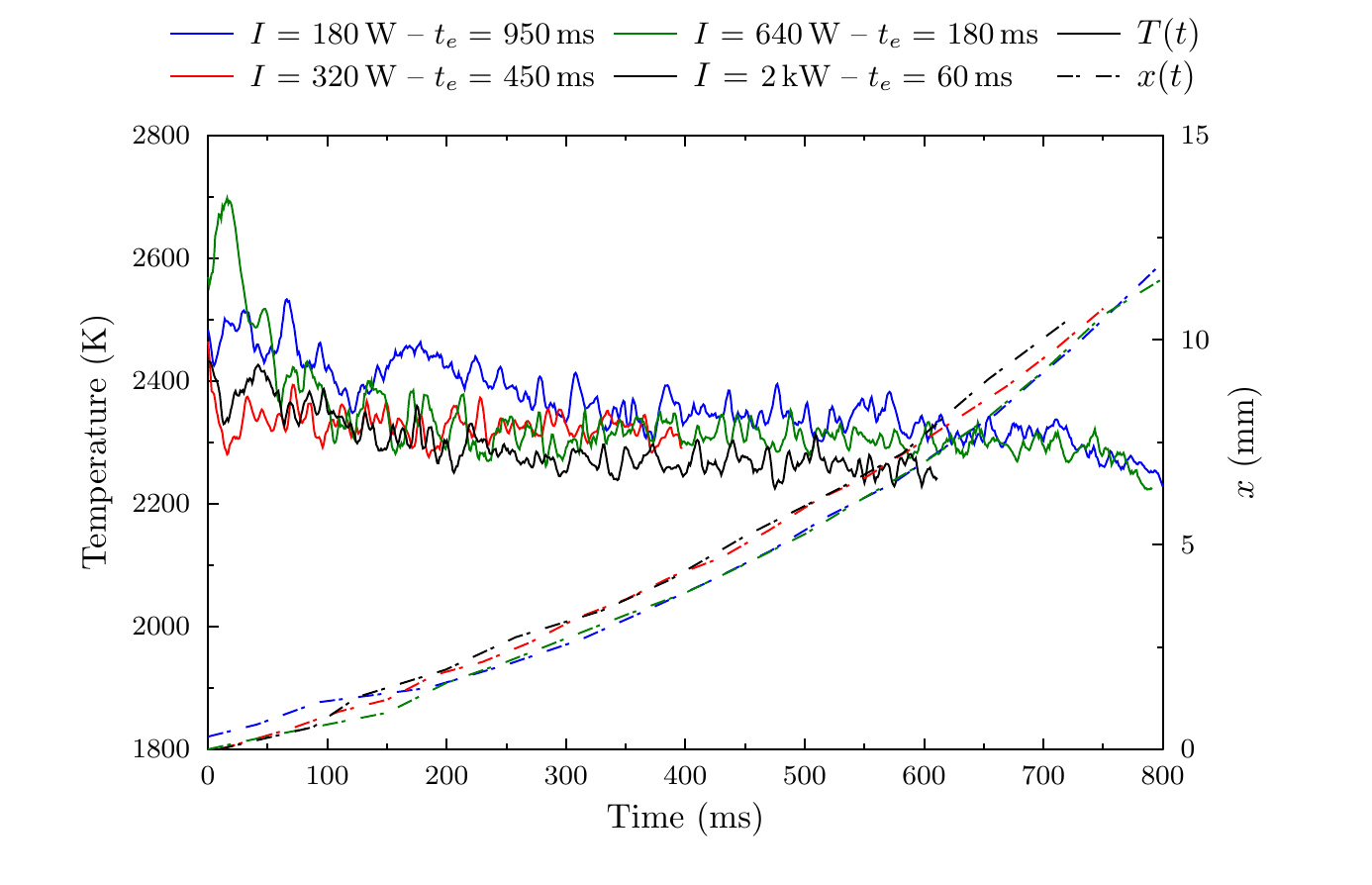}}
	\subfigure[Oxygen flow speed : 30\,m$\cdot$s$^{-1}$]{\includegraphics[width=.45\textwidth]{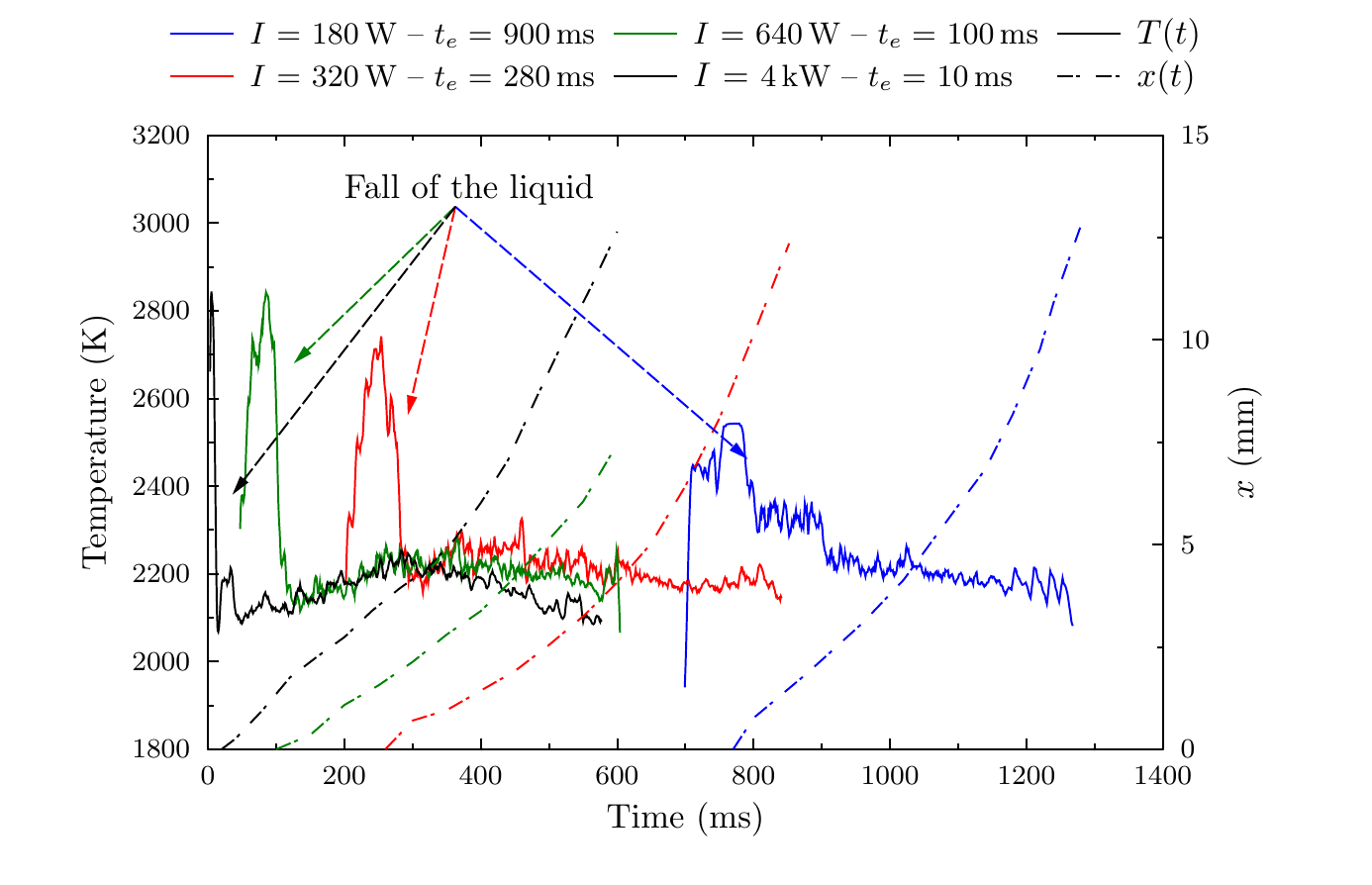}}
	\subfigure[Oxygen flow speed : 30\,m$\cdot$s$^{-1}$, shifted time]{\includegraphics[width=.45\textwidth]{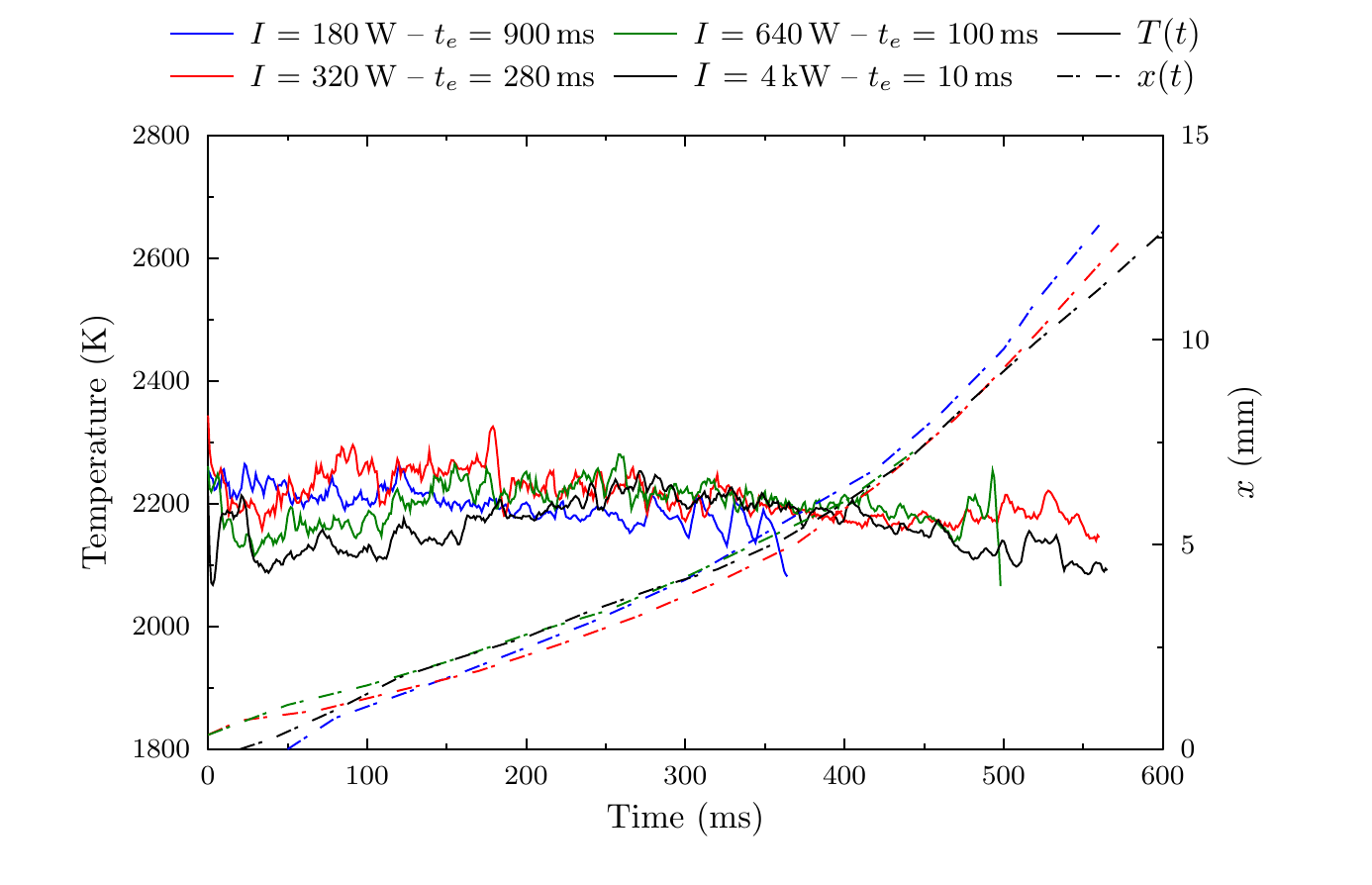}}
	\subfigure[Oxygen flow speed : 60\,m$\cdot$s$^{-1}$]{\includegraphics[width=.45\textwidth]{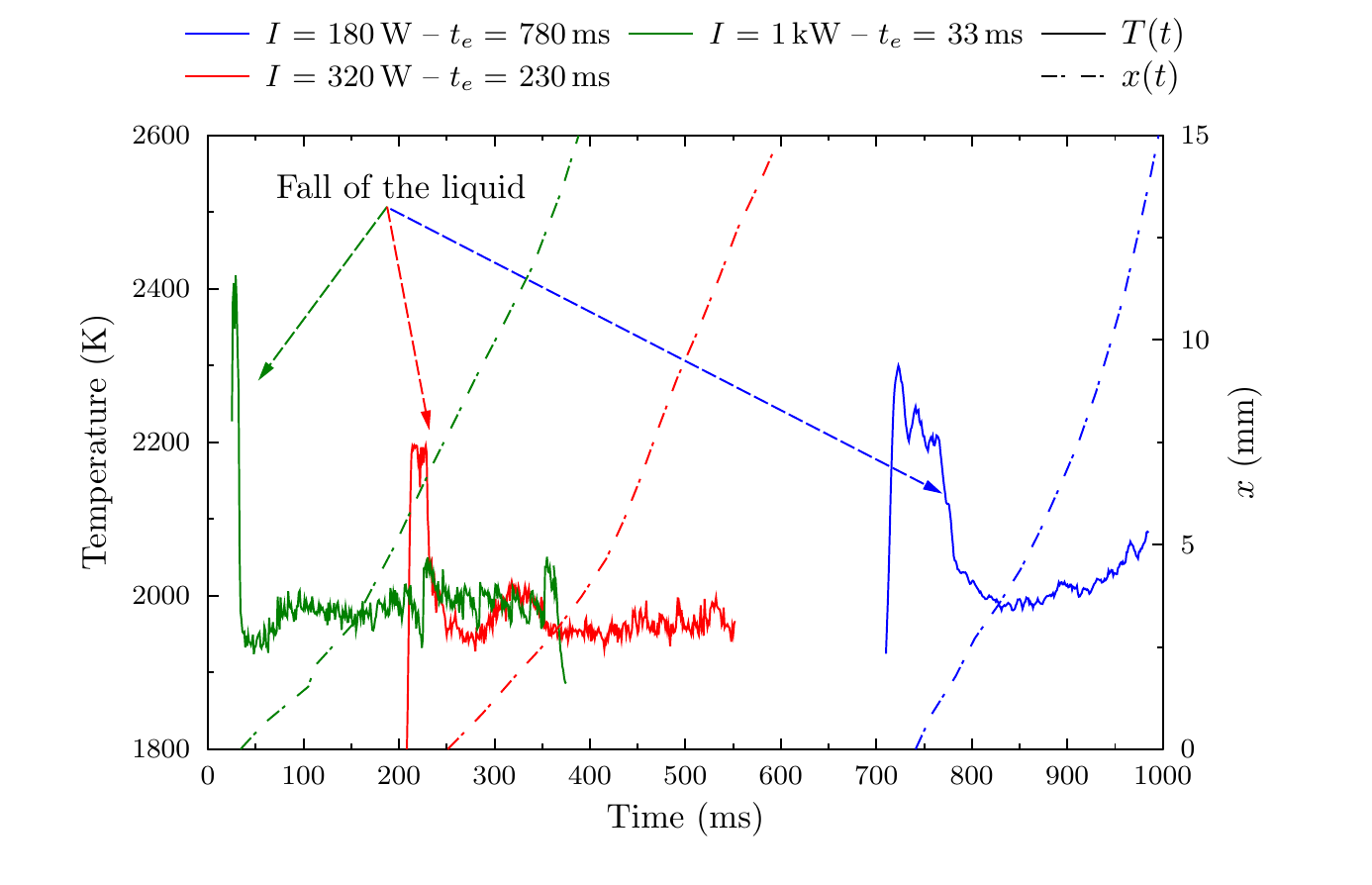}}
	\subfigure[Oxygen flow speed : 60\,m$\cdot$s$^{-1}$, shifted time]{\includegraphics[width=.45\textwidth]{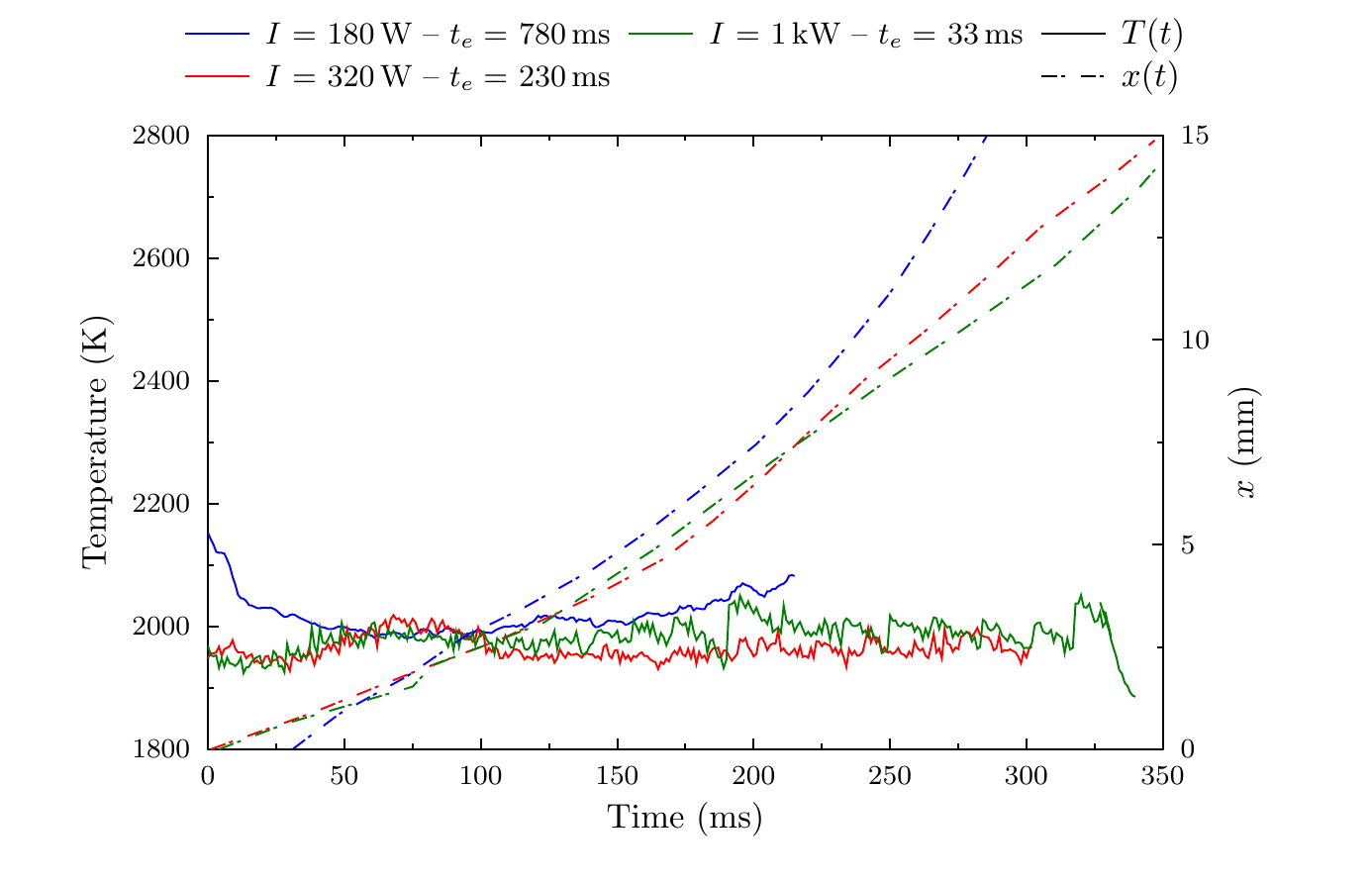}}
	\caption{Temperature and position of the rod top surface against time for different oxygen flow speeds. Figures (b)(d) and (f ) are the same as figs (a)(c) and (e), with the time $t=0$ corresponding to the fall of the liquid.}\label{fig9}
\end{figure*}

The measured temperature and position of the upper end of a mild steel rod are plotted against time in Fig.~\,\ref{fig9}. In all cases presented in this figure, the laser is switched off once the sample was ignited, so that the subsequent rod burning is self-sustained. We observe that
\begin{itemize}
	\item The onset of propagation (indicated in the graphs by ``fall of the liquid'') always corresponds to a considerable decrease of the surface temperature which varies significantly with the oxygen flow speed: whereas it is in the range $2250 \div 2500$ K for $v_0=$10\,m$\cdot$s$^{-1}$, it drops down to $1950 \div 2100$ K for $v_0=$60\,m$\cdot$s$^{-1}.$
    \item The rate of rod regression grows rapidly with the oxygen flow speed.
	\item The temperature and position curves for $v_0\geqslant$ 10\,m$\cdot$s$^{-1}$ show a good collapse upon matching the moments corresponding to the onset of dynamic combustion.
\end{itemize}	
The last observation means that for sufficiently large oxygen flow speed, the properties of dynamic combustion are independent of the preceding rod evolution, in particular, of the laser power and other specifics of ignition. The main goal of the subsequent theoretical development is to describe analytically these properties, in particular, to account for the observed increase of the metal consumption rate with the oxygen flow speed. Experimental results on the rod combustion under elevated pressures will be presented afterwards.

\section{Analytical description of the rod combustion}

\begin{figure}
	\centering
	\includegraphics[width=\textwidth]{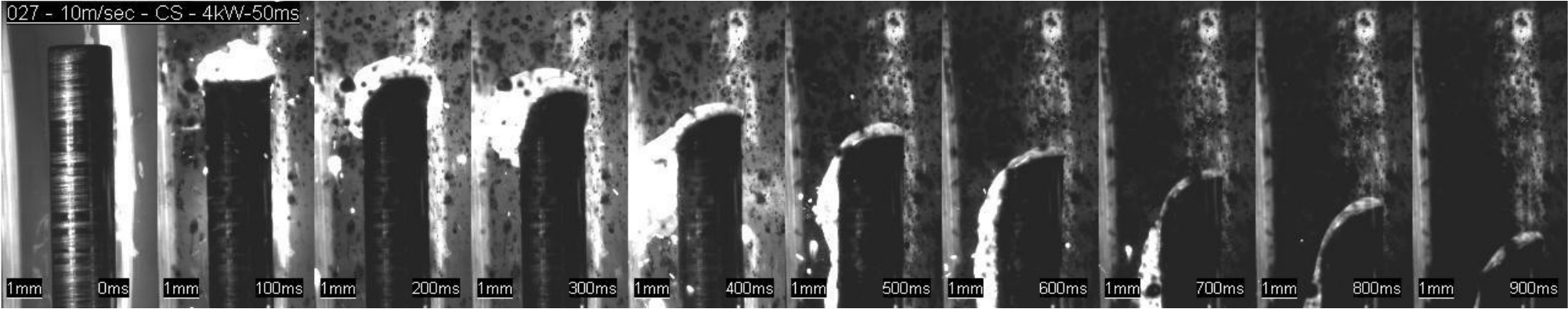}
	\caption{Regression of a carbon steel rod in an oxygen flow of 10\,m$\cdot$s$^{-1}$ speed}\label{photoseries}
\end{figure}

The series of side-view images in Fig.~\ref{photoseries} represents a typical history of the rod evolution showing its main stages: an induction period preceding the static combustion, the initial phase of which is seen on the second frame, and which ends at $t=400$\,ms with the fall of the liquid cap; it is followed by the rapid dynamic combustion characterized by a large slope of the interface between solid metal and thin liquid flow. In this natural order these stages are described analytically in Secs.~\ref{induction}--\ref{dynamic}, respectively.

\subsection{Induction stage}\label{induction}

\begin{figure}
	\centering
	\includegraphics[width=.4\textwidth]{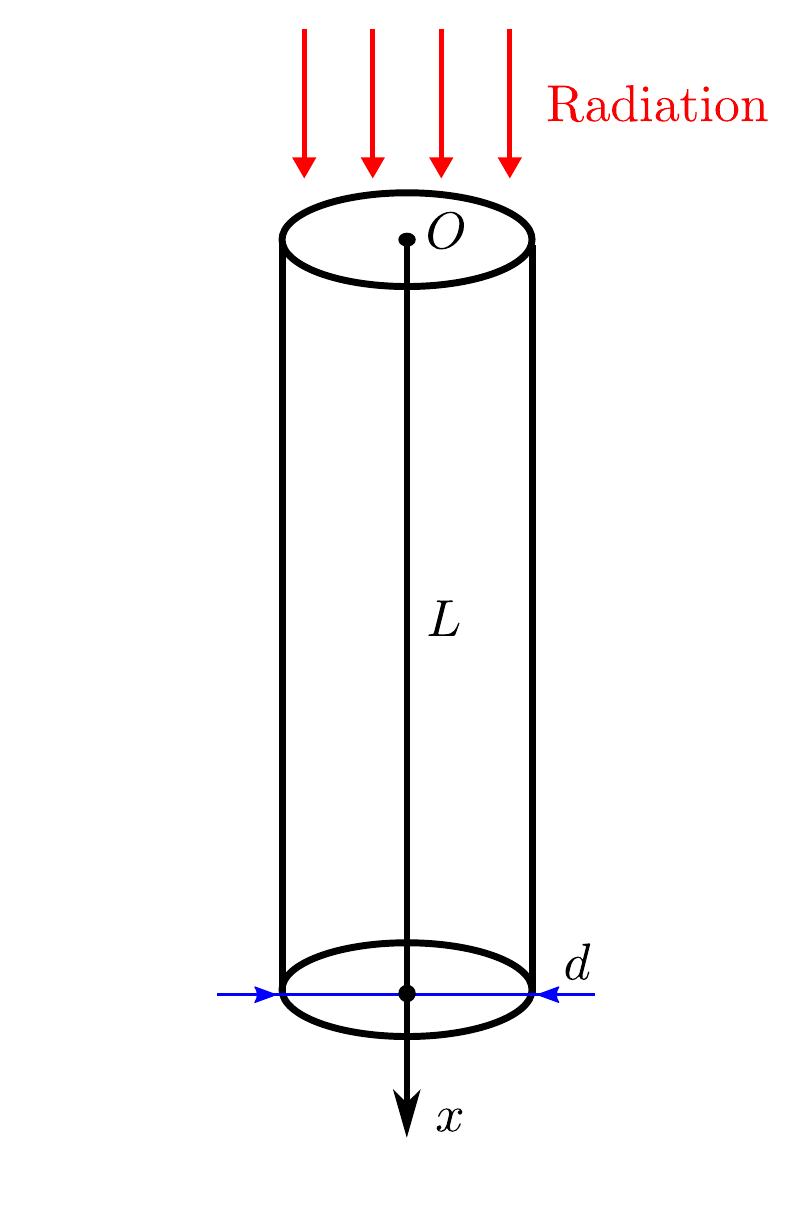}
	\caption{The laser-irradiated metal rod.}\label{rod}
\end{figure}

Given the above-described experimental conditions, we consider an iron or steel cylindrical sample placed vertically, and assume that its length, $L,$ is much larger than the diameter $d$ of its circular cross-section. Let the upper end of the sample be laser-irradiated for a time $t_e,$ the power $I$ of the electromagnetic radiation being uniformly distributed across the face surface and constant during the exposure time, Fig.~\ref{rod}. Since this surface is perpendicular to the sample axis, the radiation heat loss is negligible, and the sample is initially homogeneous (in particular, its initial temperature $T_0$ is constant), the heating process is one-dimensional. It can be described by the following equation
\begin{eqnarray}\label{fouriereq}
\frac{\partial T}{\partial t} = \chi\frac{\partial^2 T}{\partial x^2} + \alpha\delta(x), \quad \alpha = \frac{2I}{\rho c_p s}\,,
\end{eqnarray}
\noindent where $x$ is the axial co-ordinate, $\chi,\rho,$ and $c_p$ are the metal thermal diffusivity, density and specific heat, respectively, $s=\pi d^2/4,$ and the Dirac's function describes the heat release due to the irradiation of the surface $x=0.$ Assuming $\rho,c_p$ independent of temperature, the solution of this equation, satisfying given initial ($T=T_0$ at the initial instant $t=0$) and boundary conditions (absence of heat losses at the sample surface) reads
\begin{eqnarray}\label{fouriereqsol}
T(x,t) = T_0 - \frac{\alpha|x|}{2\chi} +\frac{\alpha}{4 \sqrt{\pi\chi^3 t}}\int\limits_{-\infty}^{+\infty}dx'|x'|\exp\left\{-\frac{(x-x')^2}{4\chi t}\right\}\,.
\end{eqnarray}
\noindent The integral over $x'$ is to be taken here over a domain symmetric with respect to the origin to guarantee vanishing of the heat flux at the upper end, and extending this domain over the whole infinite line implies neglecting effects related to the lower sample end.

As intensive metal oxidation ({\it i.e.}, ignition) begins once the metal is melted, the ignition time, $t_i,$ is equal to the exposure time needed to reach the melting temperature, $T_m,$ at the surface $x=0.$ Therefore, we find from Eq.~(\ref{fouriereqsol})
$$T_m-T_0 = \frac{\alpha}{4 \sqrt{\pi\chi^3 t_i}}\int\limits_{-\infty}^{+\infty}dx'|x'|\exp\left\{-\frac{x'^2}{4\chi t_i}\right\} = \alpha\sqrt{\frac{t_i}{\pi\chi}}\,,$$ which after substitution of the expression for $\alpha$ yields the induction time
\begin{eqnarray}\label{inductiontime}
t_i=\frac{\pi\chi}{4}\left[\frac{\rho c_p(T_m-T_0)s}{I}\right]^2.
\end{eqnarray}
\noindent

\begin{figure}
	\centering
	\includegraphics[width=.6\textwidth]{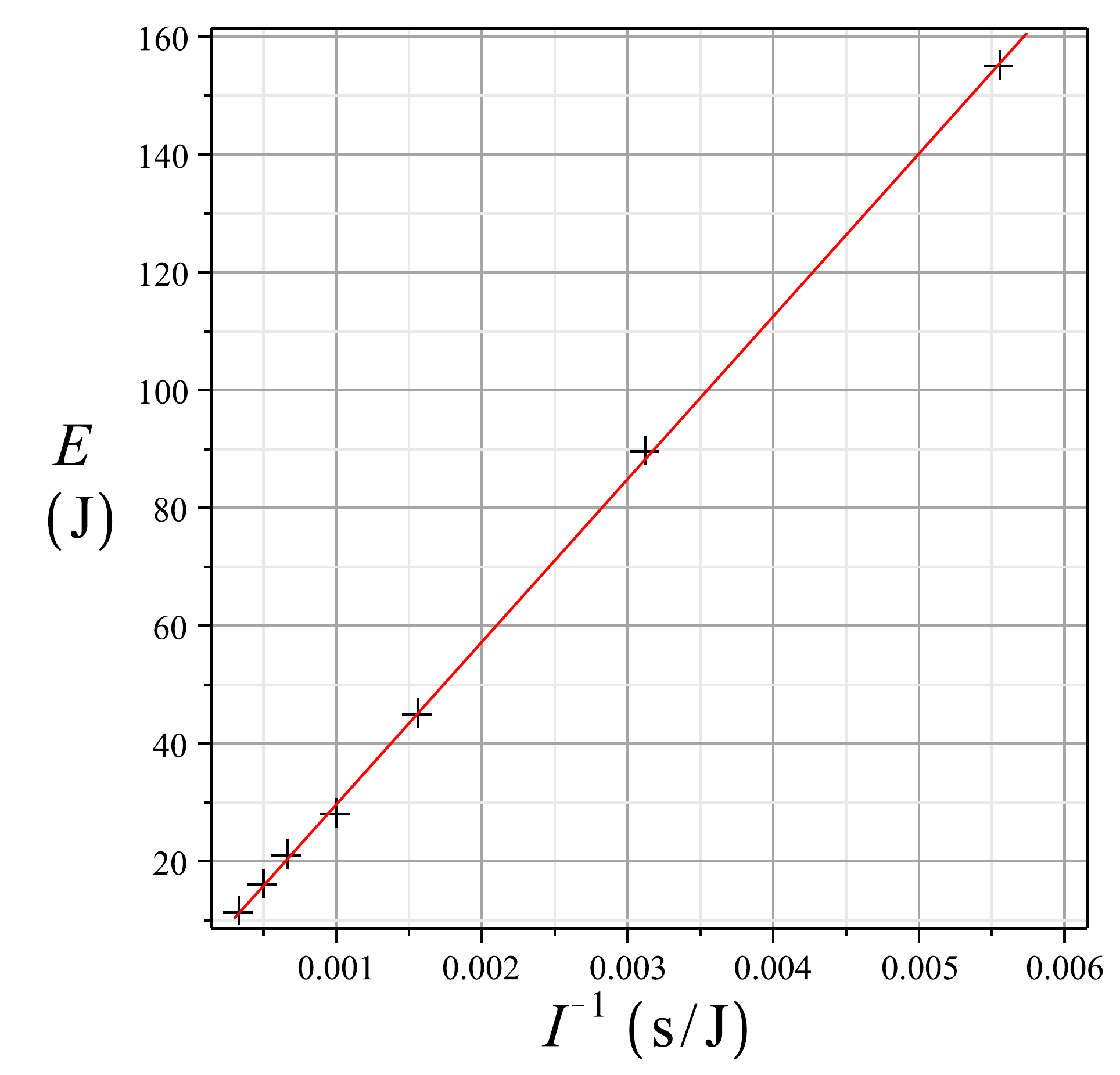}
	\caption{Total energy required to ignite a stainless steel rod of $3\,$mm diameter as a function of inverse laser power. The line represents the best linear fit, the marks -- experimental data.}\label{invpower}
\end{figure}

For the purpose of experimental verification, it is convenient to rewrite this formula by noting that it implies a linear relation between the laser power and the total energy, $E=It_i,$ transferred to the sample during the induction period, namely,
\begin{eqnarray}\label{inductionconst}
E=\frac{\beta}{I}, \quad \beta \equiv \frac{\pi\chi}{4}\left[\rho c_p(T_m-T_0)s\right]^2.
\end{eqnarray}
\noindent Figure \ref{invpower} compares this formula with the experimental data.

\subsection{Static combustion}

Subsequent to ignition is the stage of static combustion. As we have seen, during this stage liquid metal melted by the laser radiation is slowly oxidized, the energy released in the chemical reaction being spent primarily on further reheating of the liquid cap growing on the upper rod end. The heat outflow from this region to the solid part is weak despite the temperature rise, because the cap thickness is comparatively large (of the order of the rod diameter), so that the temperature gradient is relatively small. The shape of the cap and its kinematics are controlled by the surface tension and gravity. The latter affects the liquid in two ways: one is the usual bulk force acting on each liquid element, and the other is the baroclinic effect -- gravity-induced vorticity generation at the solid-liquid interface. The value of the liquid velocity curl, $\omega,$ at this interface is \cite{hayes1957}
\begin{eqnarray}\label{baroclinic}
\omega_g = \frac{\theta - 1}{\theta}\frac{g\sin\varphi}{u_n}\,,
\end{eqnarray}
\noindent where $g$ is the gravity acceleration, $u_n$ the normal velocity of the solid-liquid interface, $\varphi$ its local inclination angle with respect to the horizontal, and
$$\theta=\frac{\rho}{\rho_l}\,,$$ $\rho$ $(\rho_l)$ being the solid (liquid) density.
\begin{figure}
\centering
\includegraphics[width=.3\textwidth]{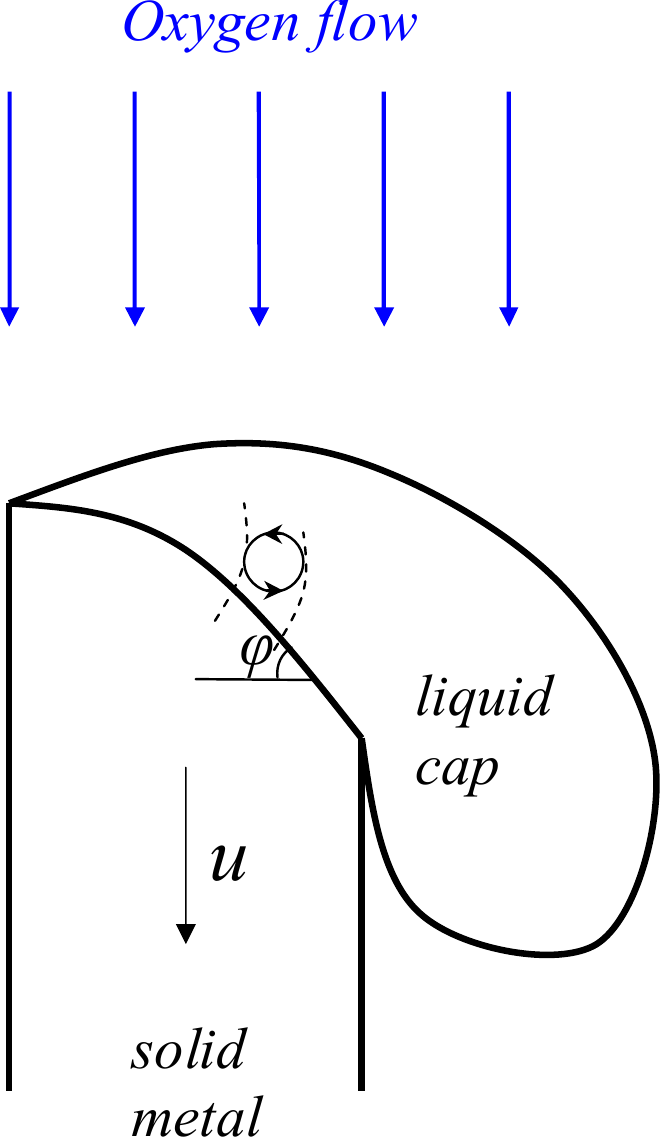}
\caption{Flow structure in the liquid cap on inclined interface. Shown are two neighboring stream lines right after they crossed the interface (broken lines), and a rotating fluid element.}\label{sketch1}
\end{figure}
If the solid-liquid interface remains horizontal ($\varphi = 0$), then $\omega_g = 0$ and nothing prevents the cap from flowing down the rod when the surface tension fails to hold the cap, that is, when its size exceeds the rod diameter by more than the capillary length (which is of the order $\sqrt{\sigma/g\rho_l},$ where $\sigma$ is the coefficient of surface tension, $\sigma\approx 2\,$N/m for iron at $T\approx 2000\,$K). Things get more complicated when the solid-liquid interface becomes inclined during the liquid cap evolution. The point is that the local rotation of the liquid elements, generated by the baroclinic effect, tends to roll them {\it up} the inclined interface (that is, the rotation is counterclockwise on the sketch shown in Fig.~\ref{sketch1}, so that its angular velocity, $\omega/2,$ is positive). Furthermore, the velocity circulation is conserved in an ideal flow (Thomson theorem), which in the simplified two-dimensional representation of Fig.~\ref{sketch1} corresponds to conservation of $\omega.$ On the other hand, the liquid flowing down along the rod has $\omega$ of the opposite sign. In a truly two-dimensional flow, therefore, the liquid cap would had to ``wait'' until this vorticity is decayed by the viscous friction, before it starts to move along an inclined surface. In the actual three-dimensional situation, however, vorticity may reverse its sign without assistance of the viscous forces, via fragmentation of the flow into vertical cells leaning over in the transverse direction, which results in a vorticity flip-over as sketched in Fig.~\ref{flip}. In any case, the cap motion is comparatively slow as it is driven by the slow ordinary oxygen diffusion and heat conduction in a laminar flow. In view of this, it can be naturally called the {\it slow onset of dynamic combustion.}

\begin{figure}
\centering
\includegraphics[width=.6\textwidth]{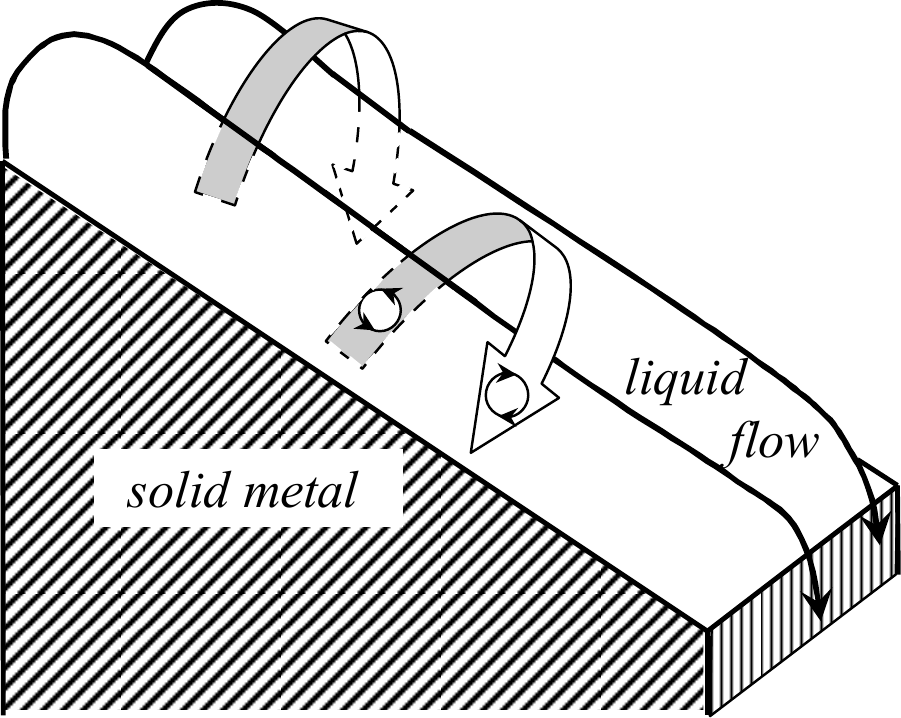}
\caption{Three-dimensional reversal of the fluid vorticity. Wide arrows show leaning over of the fluid cells, small arrows -- local fluid rotation.}\label{flip}
\end{figure}

The cap detachment becomes more rapid in the presence of a downward oxygen flow, because of the viscous drag exerted on it by the gas. Also, if the flow speed is high enough to turbulize the liquid, the oxygen consumption rate greatly increases because the oxygen transfer through the liquid flow is enhanced by the turbulent pulsations. We then have a {\it fast onset of dynamic combustion.} A quantitative criterion distinguishing the two cases is the minimal shear stress of the oxygen flow required to compensate the baroclinic vorticity, $\tau = \eta\omega_g$ ($\eta$ -- dynamic viscosity of the liquid). Writing $\tau = \rho_0 v^2_*,$ where $\rho_0$ is the oxygen density, the friction velocity $v_*$ can be estimated using the law of the wall discussed in the next section.

\subsection{The theory of dynamic combustion}\label{dynamic}

\begin{figure}
	\centering
	\includegraphics[width=.4\textwidth]{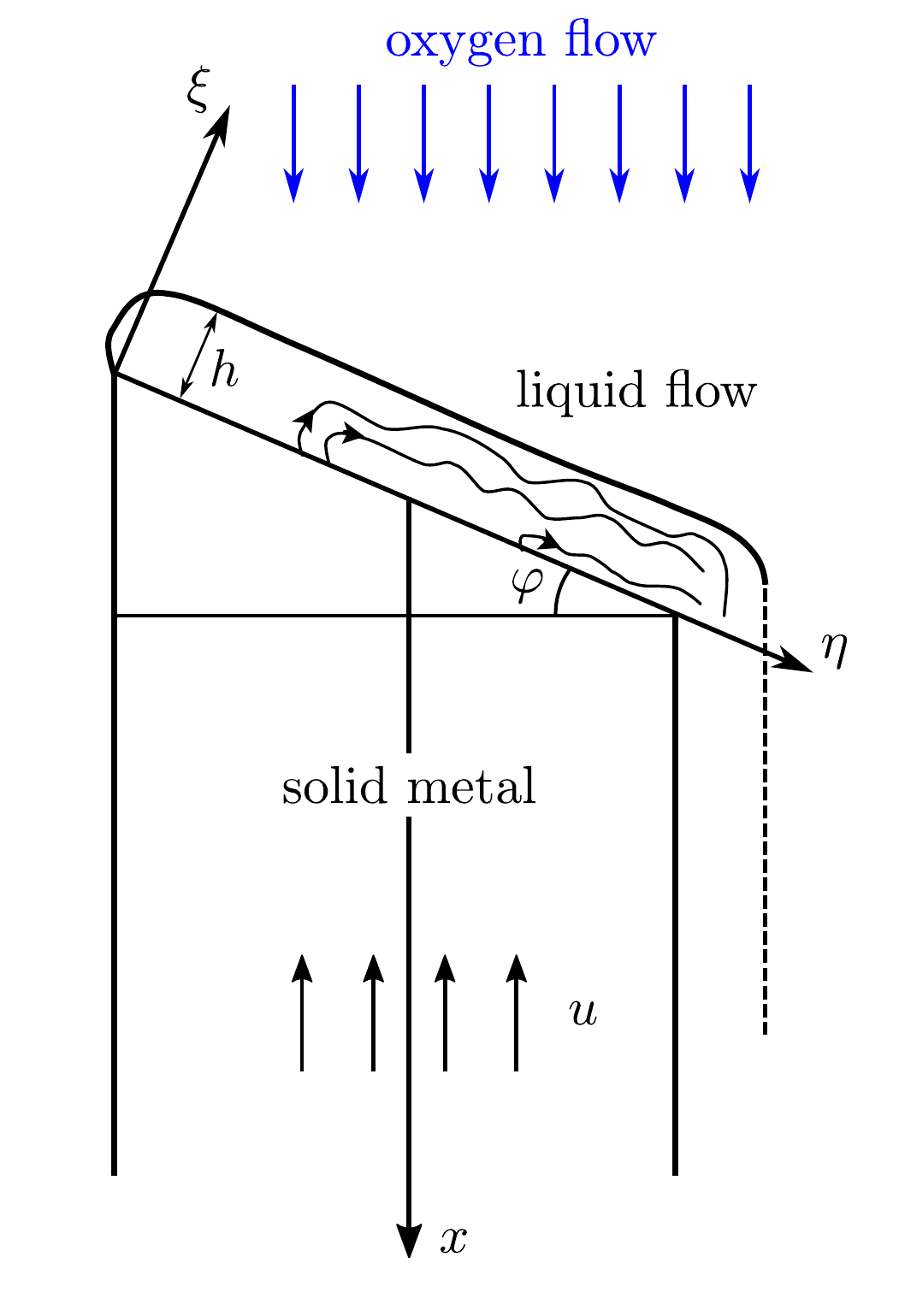}
	\caption{Schematics of the upper rod end in the regime of dynamic combustion.}\label{inclinedrod}
\end{figure}

The experimental evidence presented in Sec.~\ref{flow} is that in a sufficiently fast oxygen flow, the slow detachment of a liquid cap from the upper rod end gives way to an entirely different combustion regime, the dynamic combustion, which is characterized by a rapid consumption of the metal. The slope of the solid-liquid interface is now large, the liquid mixture of melted metal and its oxide streaming down along it as is illustrated by the right half of Fig.~\ref{photoseries} and schematized in Fig.~\ref{inclinedrod}. The vertical propagation velocity of the interface, $u,$ is a few centimeters per second, but the liquid flow velocity, $v,$ is much larger -- usually by a factor of $10$ to $10^2,$ depending primarily on the interface inclination and the oxygen flow speed. By this reason, the liquid layer thickness, $h,$ is small compared to the interface length, being normally less than $0.5\,$mm.

As this change of the burning regime is caused by the oxygen flow, it is its interaction with the liquid metal that drives the combustion process. The main characteristic of this interaction is the friction velocity $v_*$ that determines the shear stress exerted on the liquid by the flow, and which can be estimated using the law of the wall
\begin{eqnarray}\label{walllaw}
v_0 \approx 2.5v_*\ln \frac{v^2_*d}{v_0\nu_0},
\end{eqnarray}
\noindent where $\nu_0$ is the kinematic viscosity of the oxygen flow. The argument of the logarithm in this formula is the ratio of the boundary layer thickness and the viscous layer thickness, estimated as $v_*d/v_0$ and $\nu_0/v_*,$ respectively. As these are only rough estimates, accuracy of this asymptotic formula is only logarithmic, that is, it requires the logarithm be sufficiently large. [It is known that the mean velocity profile of boundary layer flows at not too large Reynolds numbers is often better approximated by a power- rather than the logarithmic law \cite{schlichting}. However, a few-percent difference between the two approximations is well within the accuracy of the subsequent applications of Eq.~(\ref{walllaw}).]

The following observation is of major importance for analytical description of the dynamic stage. The thermal diffusivity of solid iron or steel is about $0.1\,$cm$^2/$s, so that for a sample of several centimeters in length, the dimensionless ratio $uL/\chi \gg 1.$ Substituting $L = u\Delta t,$ where $\Delta t$ is the combustion time, we conclude that the dynamic stage is characterized by
\begin{eqnarray}\label{staticcond}
\frac{u^2\Delta t}{\chi} \gg 1.
\end{eqnarray}
\noindent This condition means that during most part of the combustion process the heat transfer can be treated as quasi-steady in the rest frame of the solid-liquid interface. Indeed, the heat transfer in the bulk of the sample obeys
\begin{eqnarray}\label{heattransfer}
\frac{\partial T}{\partial t} + (\bm{u}\bm{\nabla})T = \chi\Delta T.
\end{eqnarray}
\noindent If the first term on the left of this equation was larger than the second, during a time of the order of $\Delta t$ (which is the only characteristic time in the rest frame of the solid-liquid interface held at a constant temperature $T_m$) the temperature would be appreciably raised at distances $\sim \sqrt{\chi\Delta t}$ from the upper rod end. But then the ratio of the first term on the left of Eq.~(\ref{heattransfer}) to the second would be $\sqrt{\chi/u^2\Delta t}\ll 1,$ contrary to the assumption. Therefore, under condition (\ref{staticcond}) $\partial T/\partial t$ can be omitted, and the characteristic thermal length along the rod is actually $\chi/u,$ as found by comparing the second term with the right-hand side of Eq.~(\ref{heattransfer}).

\subsubsection{Thermal structure of the liquid layer}\label{thermalst}

\begin{figure}
	\centering
	\includegraphics[width=0.6\textwidth]{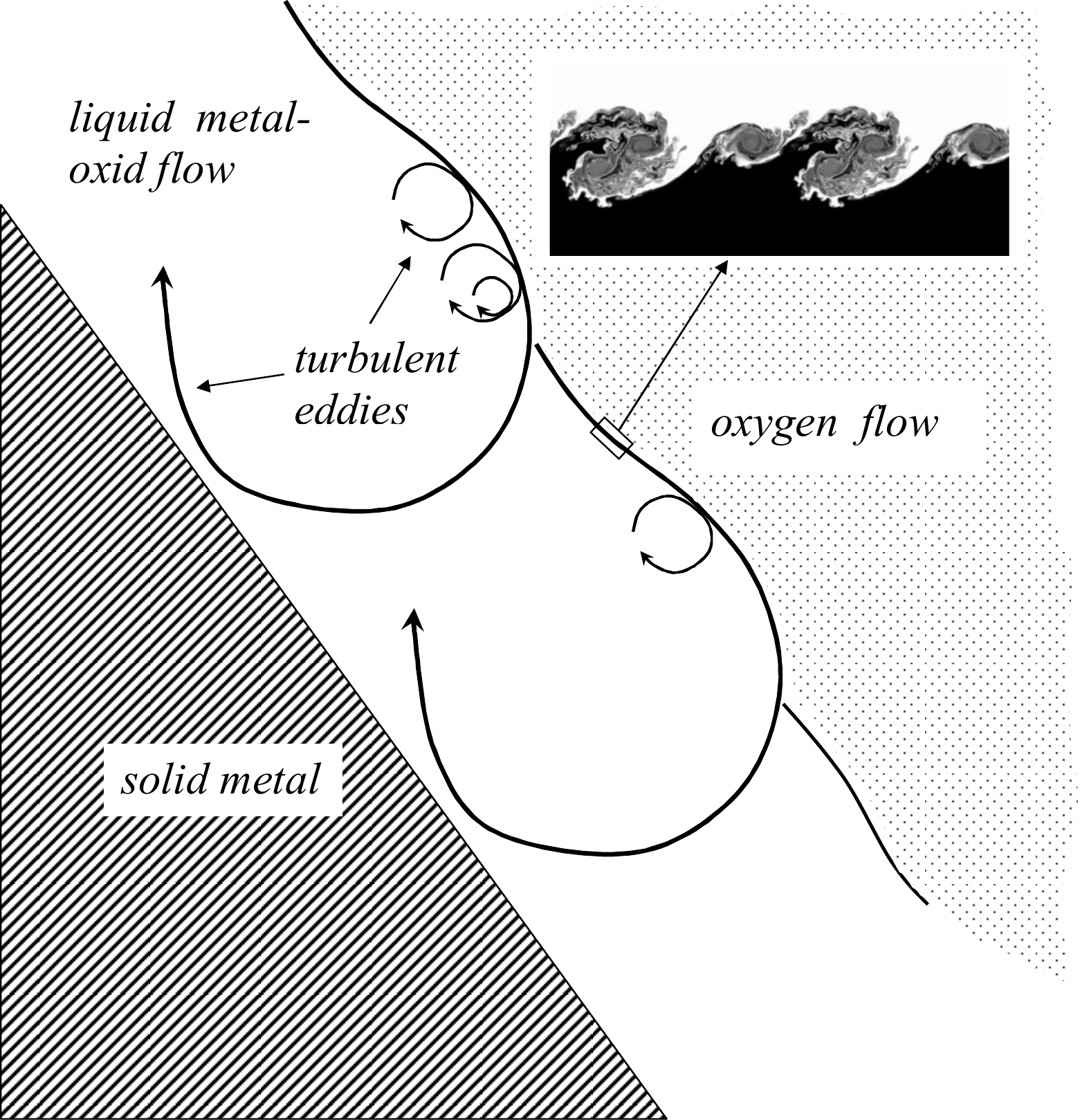}
	\caption{Schematics of the turbulent oxygen diffusion in the liquid metal in the regime of dynamic combustion.}\label{turb}
\end{figure}

In view of the smallness of kinematic viscosity of liquid iron, $\nu\approx 5\times10^{-3}\,$cm$^2/$s,\footnote{M. J. Assael, K. Kakosimos, and R. Banish, Reference data for the density and viscosity of liquid aluminum and liquid iron, Journal of physical and chemical reference data 35, 285 (2006).} and the large stream velocity, Reynolds number of the liquid flow is large,
\begin{eqnarray}\label{reynolds}
{\rm Re} = \frac{d v}{\nu}\approx 10^3-10^4,
\end{eqnarray}
\noindent implying that the flow is turbulent. An immediate important consequence is that the oxygen transport in the liquid is the turbulent diffusion. In contrast to the ordinary diffusion which under the present conditions might affect only a thin surface layer because of the smallness of the oxygen diffusion coefficient ($D \approx 10^{-4}\,$cm$^2/s$ at $T=1900\,$K), the turbulent mixing delivers oxygen into the bulk of liquid metal. This mechanism is sketched in Fig.~\ref{turb}. Turbulent motions at the liquid metal--oxygen interface enrich the metal breakers with oxygen, which is shown in the figure as a brightening of the liquid. The thickness of this oxygen-rich region is $\sim\nu/v_*,$ which is the size of the smallest eddies in the liquid (the corresponding eddies in the gas flow are of much larger size $\nu_0/v_*$). This oxygen is continuously transported from the surface layer into the bulk by slower eddies, most efficiently by the ones of the largest size $h.$

Taking into account thinness of the liquid layer, consideration can be simplified by replacing the mean oxygen concentration, $c,$ by its average value $\bar{c}$ across the layer. Also, the rod can be replaced by a bar (with a square cross-section) of the same dimension. This is possible by virtue of the turbulent nature of the liquid flow and the fact that the relation between the layer thickness and geometry of the solid-liquid interface is ultimately fixed by the overall mass conservation. We thus have the following equation for the temperature distribution in the liquid in the regime of quasi-steady dynamic combustion
\begin{eqnarray}\label{fourierliquid}
\chi\frac{\partial^2 T}{\partial \xi^2} + \frac{Q w \bar{c}}{c_p n} = 0,
\end{eqnarray}
\noindent where the $\xi$ co-ordinate is along the normal to the solid-liquid interface, $Q$ is the heat of iron combustion, $n$ is the number density of iron atoms, and $w$ is the reaction rate. The latter can also be taken constant throughout the liquid, for the following reason. In the regime under consideration, the metal oxide fraction in the liquid mixture is relatively low -- it does not exceed $20\%$ [Cf. Eq.~(\ref{fraction}) below]. The apparent activation energy of iron oxidation, $\EuScript{E},$ in this case is also comparatively small, about $30\,$kJ/mol \footnote{D. Gamburg, B. Sarychev, Trudy GIAP 7, 121 (1957).}. Therefore, the characteristic temperature interval at the iron melting point ($T_m = 1811\,$K)
$$\Delta T = \frac{RT^2_m}{\EuScript{E}} \approx 900\,{\rm K}$$ is large compared to the temperature difference across the liquid layer, which is normally $100$ to $500\,$K, so that the corresponding variation in $w$ is negligible indeed.

The boundary conditions for $T(\xi)$ are
\begin{eqnarray}\label{boundarycond}
\left.T\right|_{\xi=0} &=& T_m, \label{tempath} \\
\left.\frac{\partial T}{\partial\xi}\right|_{\xi=0} &=& \frac{u\cos\varphi}{\chi}\left(T_m -T_0 + \frac{Q_m}{c_p}\right), \label{heatfluxat0} \\
\left.\frac{\partial T}{\partial\xi}\right|_{\xi=h} &=& 0.\label{heatfluxath}
\end{eqnarray}
\noindent
Equation~(\ref{heatfluxath}) is the expression of zero heat loss from the liquid to the atmosphere, whereas Eq.~(\ref{heatfluxat0}) is obtained by integrating the steady equation
\begin{eqnarray}\label{fourier2dsteady}
-u\frac{\partial T}{\partial x} = \chi\Delta T
\end{eqnarray}
\noindent over the whole solid domain, and taking into account that $T|_{x=+\infty} = T_0;$ the temperature and its normal derivative are constant along the solid--liquid interface; and finally, the heat flux, $-\chi\rho c_p\partial T/\partial\xi,$ has a jump at this surface equal to the mass flux of metal, $\rho u\cos\varphi,$ times the heat of metal fusion, $Q_m.$

Solution of Eq.~(\ref{fourierliquid}) satisfying condition (\ref{heatfluxath}) is
\begin{eqnarray}\label{solutionfort}
T(\xi) = T_a - \frac{Qw\bar{c}}{2\chi c_pn}(\xi - h)^2, \quad T_a = {\rm const}.
\end{eqnarray}
\noindent Condition (\ref{heatfluxat0}) then yields
\begin{eqnarray}\label{relation}
\frac{u\cos\varphi}{h\chi}\left(T_m - T_0 + \frac{Q_m}{c_p}\right) = \frac{Qw\bar{c}}{\chi c_pn}\,.
\end{eqnarray}
\noindent while condition (\ref{tempath}) gives the temperature at the liquid surface, $T_a,$ -- the apparent temperature of the upper rod end
\begin{eqnarray}\label{temprod}
T_a = T_m + \frac{hu\cos\varphi}{2\chi}\left(T_m - T_0 + \frac{Q_m}{c_p}\right).
\end{eqnarray}
\noindent An important consequence of relation (\ref{relation}) is the expression for the metal oxide fraction in the liquid, $f,$ which can be found as the ratio of the iron oxidation rate, $\bar{c}whd^2/\cos\varphi,$ to the total iron consumption rate, $ud^2n.$ It follows from Eq.~(\ref{relation}) that
\begin{eqnarray}\label{fraction}
f= \frac{Q_m + c_p (T_m - T_0)}{Q}\,.
\end{eqnarray}
\noindent For iron, $Q_m = 250\,$J/g, $Q = 6000\,$J/g, $c_p = 0.45\,$J/g$\cdot$K, so that $f = 15.5\%$ for $T_0 = 300\,$K.

It is to be noted that the meaning of Eq.~(\ref{fraction}) is essentially integral, in that the value of $f$ is independent of the actual oxygen distribution across the liquid. In fact, for an arbitrary $c(\xi),$ it is readily found from Eqs.~(\ref{fourierliquid}) and (\ref{heatfluxath}) that
\begin{eqnarray}
\frac{\partial T}{\partial\xi} = \frac{Qw}{\chi c_p n}\int\limits_{\xi}^{h}d\xi'c(\xi')\,.\nonumber
\end{eqnarray}
\noindent Equation~(\ref{heatfluxat0}) then again yields Eq.~(\ref{relation}) in which now $$\bar{c} = \frac{1}{h}\int\limits_{0}^{h}d\xi c(\xi),$$ and relation (\ref{fraction}) follows as before.

\subsubsection{Structure of the liquid flow near the solid--liquid interface}

As was discussed above, the liquid metal/oxide flow is essentially turbulent. Yet, it must be laminar in immediate proximity to the solid--liquid interface because of the conditions existing thereat. This fact is important as it allows one to relate the interface slope to its propagation speed. The solid metal moves uniformly (with the speed $u$ in the rest frame of reference of the interface), while the liquid velocity at the interface is uniquely determined by the conservation of mass and momentum. Namely, the velocity component normal to the interface increases by the factor of $\theta=\rho/\rho_l,$ whereas the tangential component is continuous. In the system of Cartesian co-ordinates $(\eta,\xi)$ [see Fig.~\ref{inclinedrod}], this gives for the liquid velocity ${\bm v} = (v_{\eta},v_{\xi})$
\begin{eqnarray}\label{bcondinterface}
v_{\eta} = - u\sin\varphi, \quad v_{\xi} = \theta u\cos\varphi \quad \text{at} \quad \xi=0.
\end{eqnarray}
\noindent These relations represent boundary conditions for the turbulent flow in the bulk of the liquid. This flow is governed by the continuity and Navier-Stokes equations
\begin{eqnarray}
{\rm div}\,\bm{v} &=& 0,\label{continuity}\\
\frac{\partial\bm{v}}{\partial t} + (\bm{v}\bm{\nabla})\bm{v} &=& -\frac{1}{\rho_l}\bm{\nabla}p + {\bm g} + \nu\Delta\bm{v}.\label{navierstokes}
\end{eqnarray}
\noindent Let $\bm{\omega}$ denote the mean vorticity established, through the viscous drag, in the liquid by the external oxygen flow, $\bm{\omega} = \langle{\rm rot}\,\bm{v}\rangle.$ In the steady-on-average flow, this quantity is constant throughout the liquid layer. This follows from the conservation of velocity circulation and the fact that any non-uniformity in $\bm{\omega}$ decays due to internal friction. In particular, it has the same value near the solid--liquid interface where the flow is strictly steady and two-dimensional, and hence is described by the equations
\begin{eqnarray}\label{navierstokessteady}
\frac{\partial v_{\eta}}{\partial\eta} + \frac{\partial v_{\xi}}{\partial\xi} &=& 0, \quad
\frac{\partial v_{\xi}}{\partial\eta} - \frac{\partial v_{\eta}}{\partial\xi} = \omega,\\
v_{\eta}\frac{\partial v_{\eta}}{\partial\eta} + v_{\xi}\frac{\partial v_{\eta}}{\partial\xi} &=& -\frac{1}{\rho_l}\frac{\partial p}{\partial \eta} + g\sin\varphi + \nu\Delta v_{\eta}\,, \\
v_{\eta}\frac{\partial v_{\xi}}{\partial\eta} + v_{\xi}\frac{\partial v_{\xi}}{\partial\xi} &=& -\frac{1}{\rho_l}\frac{\partial p}{\partial \xi} - g\cos\varphi + \nu\Delta v_{\xi}\,,
\end{eqnarray}
\noindent where $\omega$ is the given constant value of $\bm{\omega} = (0,0,\omega),$ and $p$ is the liquid pressure. An obvious solution of this system reads
\begin{eqnarray}\label{navierstokessol}
v_{\eta} = - \omega\xi + {\rm const}, \quad v_{\xi} = - \frac{g\sin\varphi}{\omega}, \quad p = - \xi\rho_l g\cos\varphi.
\end{eqnarray}
\noindent It can be shown that this solution is unique under the condition of vanishing pressure gradient along the solid-liquid interface (unrestricted liquid flow). As was mentioned, it is valid only in a vicinity of this interface where the liquid flow is laminar. Now, the first of the boundary conditions (\ref{bcondinterface}) gives ${\rm const} = - u\sin\varphi,$ while the second leads to the following relation between the inclination angle, induced vorticity and the speed of rod burning
\begin{eqnarray}\label{uphirelation}
\tan\varphi = - \frac{\theta\omega u}{g}\,.
\end{eqnarray}
In application to Fig.~\ref{photoseries}, substitution of $d=0.3\,$cm, $\nu_0=0.15\,$cm$^2$/s, $v_0=10\,$m/s into Eq.~(\ref{walllaw}) gives $v_* = 120\,$cm/s, and then $\omega = - \rho_0v^2_*/\eta = - 630\,$s$^{-1}.$ Inserting this together with $\theta=1.12$ and the measured value $u = 0.8\,$cm/s into formula (\ref{uphirelation}), one finds $\tan\varphi = 0.6,$ or $\varphi\approx 30^{\,\rm o}.$

\subsubsection{Oxygen concentration in the liquid layer}

\begin{figure}
	\centering
	\includegraphics[width=0.4\textwidth]{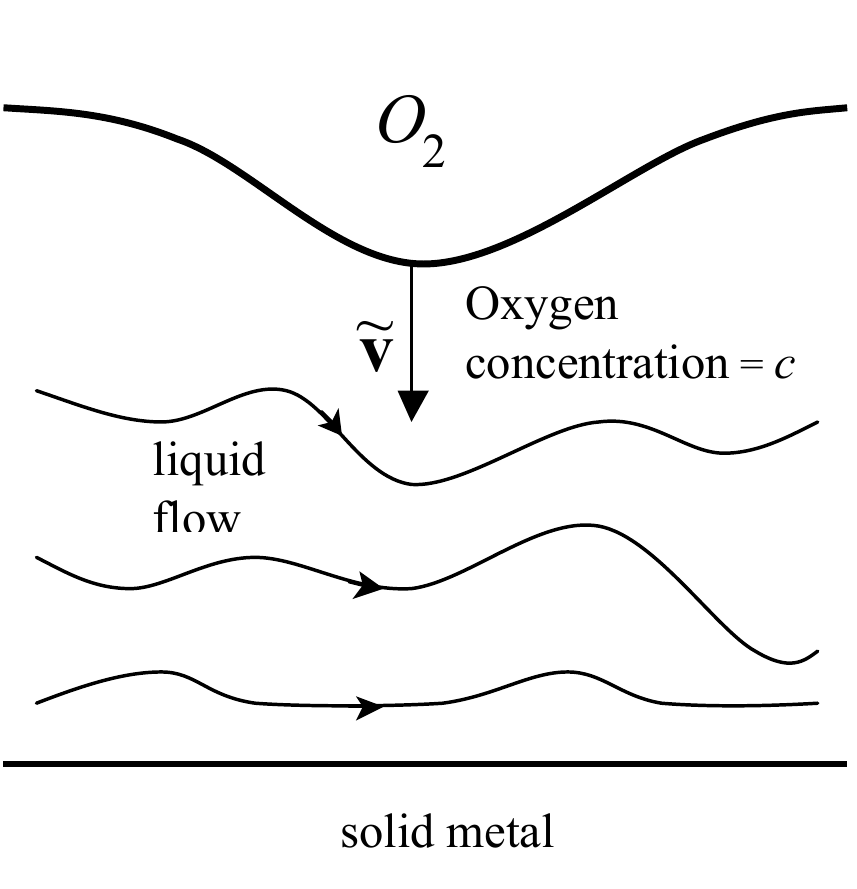}
	\caption{Schematics of enrichment of the liquid metal with oxygen used in evaluating the integral on the right of Eq.~(\ref{intc}). Shown is a fluctuation of the liquid--gas interface produced by an eddy of smallest size $\nu/v_*,$ $\tilde{\bm{v}}$ denotes a velocity fluctuation of the order $v_*.$}\label{oxygen}
\end{figure}

Oxygen transport in the liquid layer is described by the following equation for the concentration of oxygen atoms, $C,$
\begin{eqnarray}
\frac{\partial C}{\partial t} + (\bm{v}\bm{\nabla})C = D\Delta C - w C. \nonumber
\end{eqnarray}
\noindent As was already noted, the ordinary diffusion term $D\Delta C$ can be omitted as it is negligible compared to the turbulent contribution. Decomposing the fluid velocity and concentration into the mean and fluctuating parts
$$\bm{v} = \langle\bm{v}\rangle + \tilde{\bm{v}}, \quad C = c + \tilde{c}, \quad c \equiv \langle C \rangle,$$
and recalling that $c = \bar{c}$ (see Sec.~\ref{thermalst}), one finds
\begin{eqnarray}
\frac{\partial \tilde{c}}{\partial t} + (\langle\bm{v}\rangle \bm{\nabla})\tilde{c} + (\tilde{\bm{v}}\bm{\nabla})\tilde{c} = - w (\bar{c} + \tilde{c}). \nonumber
\end{eqnarray}
\noindent Statistical averaging of this equation gives
\begin{eqnarray}\label{meanc}
\langle(\tilde{\bm{v}}\bm{\nabla})\tilde{c}\rangle = - w \bar{c}.
\end{eqnarray}
\noindent According to the definition of $\bar{c},$ the left hand side is to be further averaged over the volume $V$ of the liquid layer. Rewriting the volume integral as a surface integral with the help of the flow continuity (${\rm div}\,\tilde{\bm{v}} = 0$), we obtain
\begin{eqnarray}\label{intc}
\langle(\tilde{\bm{v}}\bm{\nabla})\tilde{c}\rangle = \frac{1}{V}\int\limits_{V}dV (\tilde{\bm{v}}\bm{\nabla})\tilde{c} = \frac{1}{V}\int\limits_{S}(d\bm{s}\tilde{\bm{v}})\tilde{c},
\end{eqnarray}
\noindent
where $d\bm{s}$ is the element of the layer surface $S,$ directed outwards of the liquid. Whenever a turbulent fluctuation shifts the local surface position towards the liquid (as in Fig.~\ref{oxygen}), that is $(d\bm{s}\tilde{\bm{v}})<0,$ the fluctuation in the oxygen concentration brought in thereby equals to $(2c_0 - \bar{c}),$ while for $(d\bm{s}\tilde{\bm{v}})>0,$ the fluctuation $\tilde{c} = (\bar{c} - 2c_0).$ Here $c_0$ is the concentration of oxygen molecules in the ambient gas flow, and the factor of two accounts for the chemical adsorption of oxygen atoms. Neglecting $\bar{c}$ in comparison with $c_0,$ we thus have
$$\int\limits_{S}(d\bm{s}\tilde{\bm{v}})\tilde{c} = - 2c_0\int\limits_{S}ds |\tilde{v}_n|\,,$$ where $\tilde{v}_n$ is the normal to the surface component of fluctuating liquid velocity. By virtue of the boundary conditions at the liquid--gas interface, this is also the normal velocity component of the gas flow. Substituting this in Eq.~(\ref{intc}), writing $V = hS,$ and putting the result in Eq.~(\ref{meanc}) yields the mean oxygen concentration in the liquid
\begin{eqnarray}\label{meanco}
\bar{c} = \frac{2c_0 v_*}{wh}\,, \quad v_* = \frac{1}{S}\int\limits_{S}ds|\tilde{v}_n|\,.
\end{eqnarray}
\noindent The quantity $\frac{1}{S}\int\limits_{S}ds|\tilde{v}_n|$ appearing in this formula is denoted $v_*$ because it is the velocity fluctuation averaged over the liquid--gas interface, which is nothing but an estimate of the friction velocity.

Although the above integral calculation is a direct expression of the idea of turbulent transport, it is perhaps worth to relate it to the more conventional differential formulation. In the latter, the average $\langle \tilde{\bm{v}}\tilde{c}\rangle$ is postulated to be proportional to the mean concentration gradient, $$\langle \tilde{\bm{v}}\tilde{c}\rangle = - D_t\bm{\nabla} c,$$ $D_t$ being the coefficient of turbulent diffusion. The right hand side of Eq.~(\ref{intc}) then becomes
\begin{eqnarray}
- \frac{D_t}{V}\int\limits_{S}(d\bm{s}\bm{\nabla} c) = - \frac{D_t}{h}\left.\frac{\partial c}{\partial\xi}\right|_{\xi=h}.
\end{eqnarray}
\noindent The mean concentration increases from zero to $2c_0$ between the solid--liquid and liquid--gas interfaces. Therefore, estimating the concentration gradient by $2c_0/h,$ Eq.~(\ref{meanc}) gives
\begin{eqnarray}
\bar{c} = \frac{2c_0D_t}{wh^2},
\end{eqnarray}
\noindent and comparison with Eq.~(\ref{meanco}) yields an expression for the turbulent diffusion coefficient $$D_t = v_* h.$$ This expression is obviously in accordance with the view on the oxygen diffusion as a large-eddy transport of local fluctuations in the oxygen concentration which are caused by the turbulent velocity pulsations of the order $v_*$ at the liquid-gas interface.

\subsubsection*{Normal speed of metal burning}

Combining Eq.~(\ref{meanco}) with Eq.~(\ref{relation}) gives the following important expression for the normal speed of propagation of the solid--liquid interface with respect to the metal, $u_n = u\cos\varphi,$ that is, the local rate of metal consumption,
\begin{eqnarray}\label{normalspeed}
u_n = v_*\frac{2c_0}{fn}\,,
\end{eqnarray}
\noindent where $f$ is given by Eq.~(\ref{fraction}). Under normal pressure and temperature, $c_0 = 2.4\times 10^{19}\,$cm$^{-3},$ whereas the density of atoms in the liquid iron $n = 7.4\times 10^{22}\,$cm$^{-3},$ so that $c_0/n=0.32\times 10^{-3}$; in the case of Fig.~\ref{photoseries} ($v_0=10\,$m/s, $v_* = 120\,$cm/s), Eq.~(\ref{normalspeed}) yields $u_n = 0.54\,$cm/s.

\subsubsection*{Normal speed in the absence of oxygen flow}

Since the oxygen gas density is proportional to its pressure, $c_0\sim p_0,$ a self-sustained propagation of the reaction is possible at elevated oxygen pressures, $p_0,$ even in the absence of an oxygen flow. The propagation speed in this case is lower, as the liquid velocity fluctuations at the liquid--gas interface are now only due to the turbulent motion caused by the decaying baroclinic vorticity, $|\tilde{v}_n| \sim \omega_g h.$ Substituting this for $v_*$ in Eq.~(\ref{normalspeed}), and using Eq.~(\ref{baroclinic}) gives an estimate for the normal speed
\begin{eqnarray}
u_n \sim \frac{\theta - 1}{\theta}\frac{g\sin\varphi}{u_n}\frac{2c_0h}{fn}\,, \nonumber
\end{eqnarray}
\noindent or
\begin{eqnarray}\label{normalspeed0}
u_n \sim \sqrt{\frac{\theta - 1}{\theta}\frac{2c_0hg\sin\varphi}{fn}}\,.
\end{eqnarray}
\noindent Thus, the normal speed of self-sustained propagation $\sim p^{1/2}_0.$ It is worth recalling in this connection that the same dependence on the oxygen pressure is found when the chemical adsorption ($O_2\to O+O$) of oxygen atoms at the metal surface is the rate-determining step of the reaction \cite{hirano1983}. This circumstance might lead to confusion in experimental determination of the combustion mechanism. In our experiments, however, the normal burning speed was never less than a few millimeters per second, just as predicted by the estimate (\ref{normalspeed0}). It follows that the chemical adsorption does not slow down the burning process, though it turned out to be difficult to make an accurate experimental verification of the $p^{1/2}_0$-dependence because of the large spread of the data for vanishing oxygen flow speed.

\section{Burning in oxygen flows under elevated pressures}\label{elevatedp}

\begin{figure}
	\centering
	\includegraphics[width=.8\columnwidth]{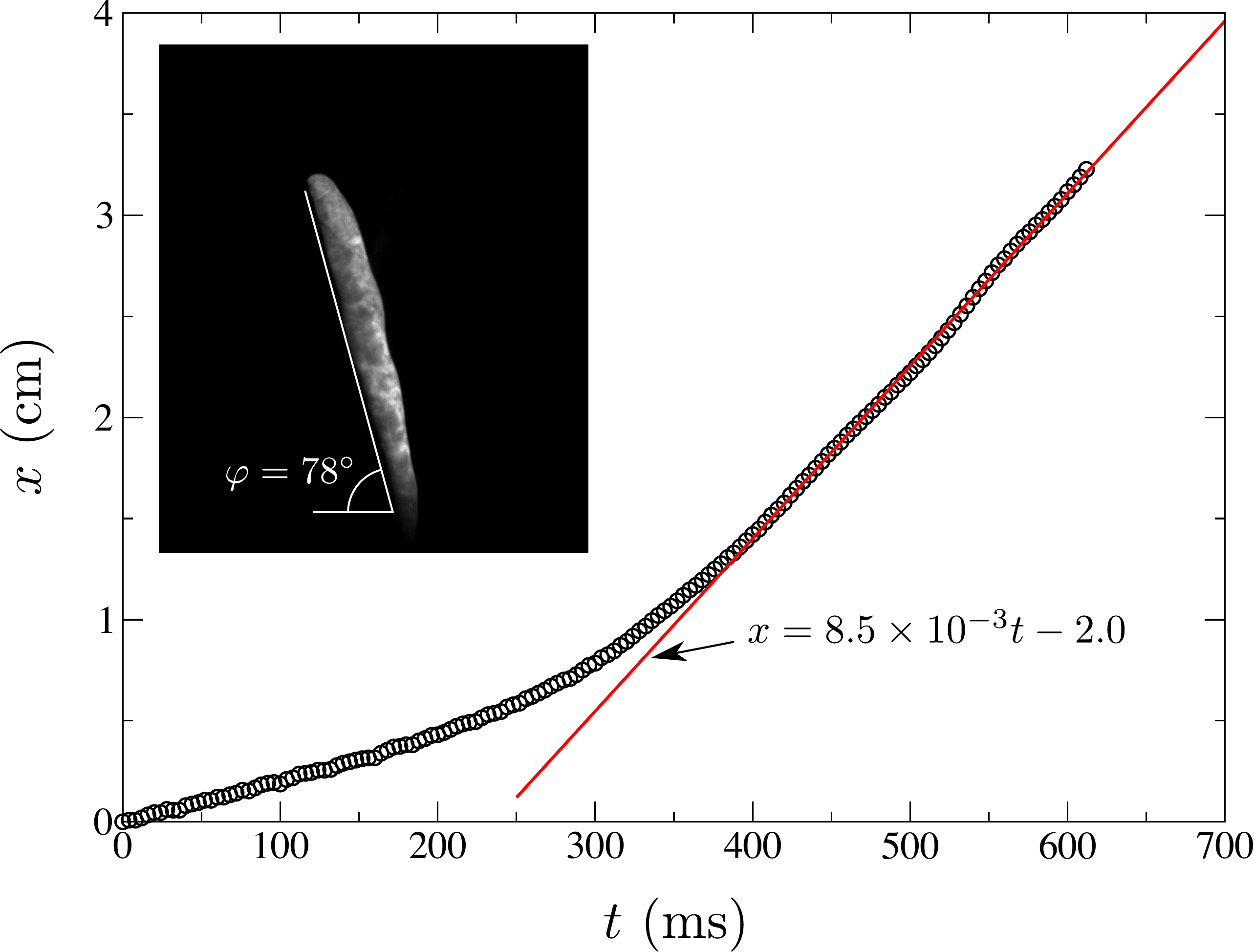}
	\caption{Measured position of the upper rod end (marks) and its linear approximation (solid line) for oxygen speed and pressure of 3\,m$\cdot$s$^{-1}$ and 8.5\,atm, respectively. Inset: determination of the solid--liquid interface slope.}\label{slope}
\end{figure}

Among the factors appearing in the expression (\ref{normalspeed}) for the normal burning speed, dependence on the oxygen concentration appears to be easiest for experimental verification. However, our experimental setup does not allow appreciable pressure changes at the oxygen flow speeds higher than $10$\,m$\cdot$s$^{-1}.$ Also, since we are not able to control azimuth of the liquid metal flow, only in a few instances it was possible to obtain a profile view of the metal flow that would allow determination of its slope and propagation speed. These are collected in Table~\ref{table}, and Figure \ref{slope} gives an example of these determinations. Experimental precision of each slope or speed determination is about 1\%--2\%, but in the absence of statistics for a given pair $v_0, p_0$ due to the reason just mentioned, it is difficult to give an estimate of the total experimental uncertainty. As to the calculational error, it is primarily due to the error in evaluating the friction velocity. It can be roughly estimated by noting that if one admits an error as large as a factor of two in the boundary layer thickness of the oxygen flow, then the relative error in $v_*$ for, say, $7$ m$\cdot$s$^{-1}$ flow speed is $\ln 2/ \ln(v^2_*d/v_0\nu_0) \approx 20\%.$ As is seen from Table~\ref{table}, the calculated and measured values for the normal burning speed agree within this accuracy.
	
\begin{table}
	\centering
	\begin{tabular}{ccccccc}
	  \toprule[0.8pt]
		$v_0$ (m$\cdot$s$^{-1}$) & \text{$p_0$ (atm)} & $v_*$ (cm$\cdot$s$^{-1}$) & $u^{\mathrm{exp}}$(cm$\cdot$s$^{-1}$) & 		$\varphi^{\mathrm{exp}}(\mathrm{deg})$ & $u_n^{\mathrm{exp}}$ (cm$\cdot$s$^{-1}$) & $u_n$ (cm$\cdot$s$^{-1}$) \\
	  	\midrule
	  	3 & 8.5  & 45.6 & 8.5 & 78 & 1.8 & 1.7 \\
	  	3 & 11.0 & 45.6 & 12.7  & 79 & 2.4 & 2.1 \\
	  	\midrule
	  	7 & 6.0 & 89.5 & 14.6  & 80 & 2.4 & 2.3 \\
	  	7 & 8.5   & 89.5 & 21.3 & 81 & 3.4 & 3.2 \\
	  	7 & 11.0  & 89.5 & 19.0 & 77 & 4.3 & 4.2 \\
	  	\bottomrule[0.8pt]
	\end{tabular}
	\caption{Comparison of the calculated ($u_n$) and measured ($u_n^{\mathrm{exp}}$) normal propagation speeds of metal burning at various oxygen flow speeds and pressures.}\label{table}
\end{table}

\section{Discussion and conclusions}

The results of our study show that the process of iron and mild steel combustion is strongly affected by the motions of ambient oxygen atmosphere: it turns out to be driven by an axial oxygen flow already at the flow speeds of a few meters per second. The main reason for this dramatic change in the combustion regime as compared to the burning in a still atmosphere is the changeover of the oxygen transport through the liquid metal from ordinary to turbulent diffusion. The small-scale turbulent motions at the liquid--gas interface enrich the melt with oxygen which is then rapidly redistributed over the whole liquid by larger eddies. The rate of metal consumption is thus controlled by the rate of this enrichment, which in turn is determined by the oxygen concentration and its speed. This is quantified by Eq.~(\ref{normalspeed}) for the normal speed of the solid--liquid interface: it is proportional to the oxygen concentration, and up to a logarithmic factor, to its speed. It is also naturally in inverse proportionality to the fraction of the metal oxide: the larger this fraction, the lesser rate of metal melting is required to sustain a steady regime of metal combustion. It is worth to stress in this connection that the process of oxygen enrichment by the turbulent motions at the liquid--gas interface is autonomous in that it is independent of the metal consumption rate. It is by this reason that it turned out to be possible to express the normal burning speed as a function of the oxygen concentration and its flow speed. This situation resembles that found in premixed flames with high activation energy of the reaction, in which case the reaction zone within the flame front is well-known to acquire similar independence of the flame propagation speed.

Formula (\ref{fraction}) for the oxide fraction itself deserves further discussion. The point is that this formula does not involve the speed $u$ or any other dynamical characteristic of the sample burning. This suggests that its validity is not limited by the quasi-steady condition. Indeed, it is merely an expression of the fact that the front of combustion -- the solid--liquid interface -- propagates at constant temperature (the metal melting point), and Eq.~(\ref{fraction}) just states that $(fQ)$ is the heat needed to preheat and melt a unit mass of metal. If, for some reason, the actual fraction would exceed $f,$ an increased heat flux to the solid region would increase the speed of the interface advancement, bringing more metal into the liquid flow and reducing thereby the oxide fraction.

The quantitative agreement of the calculated and measured normal burning speeds allows us to conclude that the theoretical approach developed in Sec.~\ref{dynamic} adequately describes the physics of iron combustion in oxygen flows.

\begin{acknowledgments}
K.~Kazakov is grateful to the French Government for supporting his stay at Poitiers University during the fall of 2012 under the program ``Investissements d'Avenir'' (LABEX INTERACTIFS, reference ANR-11-LABX-0017-01). The authors also thank Prof. R.~Fabbro for discussions.
\end{acknowledgments}

\end{document}